\documentclass[12pt]{wiley-article}

\usepackage[super]{natbib}
\usepackage{siunitx}
\usepackage[labelfont=bf]{caption}

\usepackage{amsmath,amsfonts,amssymb,amsthm,booktabs,color,epsfig,graphicx,hyperref,url}
\usepackage{bm}
\usepackage{mathtools}
\usepackage{soul}
\usepackage{natbib}
\usepackage{array}
\usepackage{multirow}

\newlength{\wdth}

\theoremstyle{plain}
\newtheorem{theorem}{Theorem}

\newtheorem{corollary}{Corollary}

\papertype{Original Article}

\title {Power and sample size calculation for multivariate longitudinal trials using the longitudinal rank sum test}

\author[1\authfn{1}]{Dhrubajyoti ~Ghosh PhD}
\author[1]{Xiaoming Xu PhD}
\author[1\authfn{1}]{Sheng ~Luo PhD}
\author[$\dagger$]{for the CPP Integrated Parkinson's Database}

\affil[1]{Department of Biostatistics and Bioinformatics, Duke University, Durham, NC, 27606, USA}

\corraddress{Sheng Luo, PhD, Department of Biostatistics and Bioinformatics, Duke University, Durham, NC, 27606, USA}
\corremail{sheng.luo@duke.edu}

\fundinginfo{National Institute on Aging, Grant/Award Numbers: R01AG064803 and P30AG072958.}

\runningauthor{Ghosh et al.}

\begin{document}

\begin{frontmatter}
\maketitle

\begin{abstract}
Neurodegenerative diseases such as Alzheimer's and Parkinson's often exhibit complex, multivariate longitudinal outcomes that require advanced statistical methods to comprehensively evaluate treatment efficacy. The Longitudinal Rank Sum Test (LRST) offers a nonparametric framework to assess global treatment effects across multiple longitudinal endpoints without requiring multiplicity corrections. This study develops a robust methodology for power and sample size estimation specific to the LRST, integrating theoretical derivations, asymptotic properties, and practical estimation techniques. Validation through numerical simulations demonstrates the accuracy of the proposed methods, while real-world applications to clinical trials in Alzheimer's and Parkinson's disease highlight their practical significance. This framework facilitates the design of efficient, well-powered trials, advancing the evaluation of treatments for complex diseases with multivariate longitudinal outcomes.

\end{abstract}

\let\thefootnote\relax\footnotetext{$^\dagger$Data used in the preparation of this article were obtained from the CPP Integrated Parkinson's Database assisted with the design and implementation of the data platform and/or provided data, but did not participate in the analysis of the data or writing of this report.}

\end{frontmatter}

\section{Introduction\label{sec:1}}

Neurodegenerative diseases such as Alzheimer’s disease (AD) and Parkinson’s disease (PD) affect multiple domains of functioning, including cognitive, motor, and behavioral abilities. These diseases often exhibit complex and heterogeneous progression patterns across various health outcomes over time. Some outcomes may decline significantly, while others remain stable, complicating efforts to comprehensively assess disease progression. Randomized clinical trials (RCTs) are critical for evaluating treatment efficacy in these conditions, collecting data on multiple health outcomes at repeated time points. For instance, the Bapi 301 and 302 studies \cite{salloway2014two} were Phase 3 RCTs assessing the efficacy of Bapineuzumab (Bapi) versus placebo in patients with mild-to-moderate AD. These trials collected longitudinal data on key outcomes such as the Alzheimer’s Disease Assessment Scale–Cognitive Subscale (ADAS-Cog) and the Disability Assessment for Dementia (DAD), providing a detailed view of cognitive and functional changes over time.

Accurately evaluating treatment efficacy in such trials requires statistical methods that can simultaneously account for both the temporal evolution of outcomes and their multivariate structure. This involves incorporating variability within and between outcomes as well as the correlations among them. Traditional approaches often reduce longitudinal data to cross-sectional summaries, such as changes from baseline to the last observation, applying parametric \cite{bartholomew1961test, perlman1969one, beutner2009order, lu2013halfline} or nonparametric \cite{o1984procedures, akritas1997unified, brunner2002multivariate, brunner2017rank, liu2010rank} procedures. However, these methods may fail to capture dynamic patterns of change over time, potentially overlooking treatment effects that manifest throughout the study period. Rank-based nonparametric methods for repeated measures \cite{konietschke2010testing, umlauft2019wild, rubarth2021estimation, rubarth2022ranking, brunner2002multivariate} have also been proposed, but they are typically limited to single outcomes or fail to integrate information effectively across multiple outcomes, restricting their ability to provide a comprehensive assessment of treatment effects.

To address these challenges, Xu et al. \cite{xu2025SBR} introduced the Longitudinal Rank Sum Test (LRST), a rank-based nonparametric global test derived from U-statistics to detect treatment efficacy in clinical trials with multiple longitudinal outcomes. The LRST provides a novel framework for obtaining a global assessment of treatment efficacy by aggregating temporal patterns across all outcomes into a single test statistic. This approach eliminates the need for multiplicity correction while fully leveraging the multivariate and longitudinal structure of the data, enabling researchers to evaluate the overall impact of a treatment across multiple domains. By deriving the null distribution of the test statistic, the method facilitates efficient hypothesis testing and offers a unified measure of treatment efficacy that captures both inter- and intra-outcome variability. These features make the LRST particularly valuable for trials in neurodegenerative diseases, where progression involves a combination of cognitive, motor, and behavioral outcomes.

In this article, we propose a comprehensive framework for power and sample size calculations tailored to the LRST. Our contributions include: (1) deriving the asymptotic distribution of the LRST under both null and alternative hypotheses, (2) developing closed-form expressions for power and sample size determination, (3) providing consistent estimates of power along with it's asymptotic distribution, and (4) validating these theoretical results through extensive simulation studies and real-world clinical trial applications. By integrating theoretical derivations with practical estimation techniques, we provide a robust approach for designing clinical trials that assess complex temporal outcomes, such as those observed in AD and PD.

The remainder of this article is organized as follows. Section \ref{sec:2} details the methods, with Section \ref{subsec:21} providing a brief overview of the LRST, Section \ref{subsec:22} describing the power and sample size calculation framework, and Section \ref{subsec:23} presenting power estimation and asymptotic properties. Section \ref{sec:sim} validates the methodology through numerical studies, while Section \ref{sec:realData} demonstrates real-world applications using clinical trial data. Finally, Section \ref{dec:conclusion} concludes with a discussion of implications and future research directions.

\section{Methods\label{sec:2}}

\subsection{Overview of the longitudinal rank sum test\label{subsec:21}}

In AD RCTs involving multiple longitudinal outcomes across two parallel arms (treatment vs. control or placebo), the data structure is typically represented as $(x_{itk}, y_{jtk}, t)$. Here, $x_{itk}$ denotes the change in outcome $k$ ($k = 1, \ldots, K$, where $K$ is the total number of outcomes) from baseline to time $t$ ($t = 1, \ldots, T$, where $T$ is the number of post-baseline visits) for subject $i$ ($i = 1, \ldots, n_x$) in the control group, while $y_{jtk}$ represents the corresponding change for subject $j$ ($j = 1, \ldots, n_y$) in the treatment group. The control and treatment group sample sizes are denoted by $n_x$ and $n_y$, respectively, with the total sample size $N = n_x + n_y$. 

Each outcome $k$ at time $t$ is ranked across both groups, producing mid-ranks $R_{xitk}$ for the control group and $R_{yjtk}$ for the treatment group. To summarize rank-based treatment effects, we define the average mid-ranks across all subjects, outcomes, and time points as: $\bar{R}_{x\cdot\cdot\cdot} = \frac{1}{n_xTK} \sum_{i=1}^{n_x} \sum_{t=1}^{T} \sum_{k=1}^{K} R_{xitk}$ and $\bar{R}_{y\cdot\cdot\cdot} = \frac{1}{n_yTK} \sum_{j=1}^{n_y} \sum_{t=1}^{T} \sum_{k=1}^{K} R_{yjtk}$. These terms represent the average mid-ranks for the control and treatment groups, respectively, over all measured time points and outcomes. The rank difference vector is then defined as: $
\bm{R} = \left(\bar{R}_{y \cdot 1 \cdot} - \bar{R}_{x \cdot 1 \cdot}, \ldots, \bar{R}_{y \cdot T \cdot} - \bar{R}_{x \cdot T \cdot} \right)^\intercal$, which captures the time-dependent rank-based treatment effect.

\subsubsection{Relative treatment effect and hypothesis testing}

The Longitudinal Rank Sum Test (LRST) extends the framework of Brunner et al. \cite{Brunner02book} to quantify treatment efficacy. The relative treatment effect for outcome $k$ at time $t$ is defined as: $\theta_{tk} = P(X_{tk} < Y_{tk}) - P(X_{tk} > Y_{tk})$, where $X_{tk}$ and $Y_{tk}$ are random variables corresponding to $x_{tk}$ and $y_{tk}$, with marginal distributions $F_{tk}$ and $G_{tk}$, respectively. The relative treatment effect across all outcomes at time $t$ is given by: $\theta_{t} = \frac{1}{K} \sum_{k=1}^{K} \theta_{tk}$. Finally, the overall treatment effect across all outcomes and time points is: $\bar{\theta} = \frac{1}{T} \sum_{t=1}^{T} \theta_{t}$. A positive $\bar{\theta}$ suggests that the treatment improves or slows the progression of some outcomes, leading to the hypothesis test: $H_0: \bar{\theta} = 0 \quad \text{vs.} \quad H_1: \bar{\theta} > 0$. To test this hypothesis, Xu \textit{et al}. \cite{xu2025SBR} proposed the LRST statistic:
\begin{equation} \label{eq:tlrst}
    T_{LRST} = \frac{\bar{R}_{y\cdot\cdot\cdot} - \bar{R}_{x\cdot\cdot\cdot}}{\sqrt{\widehat{var}(\bar{R}_{y\cdot\cdot\cdot} - \bar{R}_{x\cdot\cdot\cdot})}}
\end{equation}
where $\bar{R}_{y\cdot\cdot\cdot}$ and $\bar{R}_{x\cdot\cdot\cdot}$ are the average ranks across all outcomes and time points for the treatment and control groups, respectively. Larger values of the rank difference $(\bar{R}_{y\cdot\cdot\cdot} - \bar{R}_{x\cdot\cdot\cdot})$ indicate greater treatment efficacy. The denominator,  $\widehat{var}(\bar{R}_{y\cdot\cdot\cdot} - \bar{R}_{x\cdot\cdot\cdot})$, is a consistent estimator of the rank difference variance \citep{ghosh2024lrst}.

The LRST, a rank-based test based on U-statistics, leverages all available longitudinal data to provide a robust, nonparametric evaluation of treatment efficacy. If $H_0$ is rejected, the treatment is inferred to be more effective than the control. Further analysis, such as linear mixed models (LMMs), can then be used to quantify effects on individual outcomes. For further theoretical properties of the LRST, see Xu et al. \cite{xu2025SBR}.

\subsubsection{Asymptotic properties of the rank difference vector}
Under large-sample conditions, the rank difference vector $\bm{R} = \left(\bar{R}_{y \cdot 1 \cdot} - \bar{R}_{x \cdot 1 \cdot}, \ldots, \bar{R}_{y \cdot T \cdot} - \bar{R}_{x \cdot T \cdot} \right)^\intercal$ follows an asymptotic normal distribution \cite{xu2025SBR}: \[
\frac{1}{\sqrt{N}}\bm{R} \sim N\left(\frac{\sqrt{N}}{2} \pmb{\theta}, \bm{\Sigma}\right),
\]
where $\pmb{\theta} = \left(\theta_1, \ldots, \theta_T \right)^\intercal$ is the vector of relative treatment effects over time, and $\bm{\Sigma}$ is the variance-covariance matrix of the rank differences. The variance-covariance matrix $\bm{\Sigma}$ is given by:
\[
\Sigma_{t_1t_2} \sim \frac{1}{K^2} \sum_{k_1=1}^K \sum_{k_2=1}^K \left[\left(1 + \frac{1}{\lambda}\right) c_{t_1k_1t_2k_2} + (1+\lambda)d_{t_1k_1t_2k_2}\right], \quad \left(\frac{n_x}{n_y} \rightarrow \lambda \right).
\]

The terms $c_{t_1k_1t_2k_2} = Cov\left(G_{t_1k_1}(X_{t_1k_1}), G_{t_2k_2}(X_{t_2k_2})\right)$ and $d_{t_1k_1t_2k_2} = Cov\left(F_{t_1k_1}(Y_{t_1k_1}), F_{t_2k_2}(Y_{t_2k_2})\right)$ represent the covariances of the rank-based empirical cumulative distribution functions (CDFs) for the control and treatment group outcomes, respectively, across time points $t_1, t_2$ and outcomes $k_1, k_2$. Here, $G_{t_1k_1}(x) = P(X_{t_1k_1} \leq x)$ and $F_{t_1k_1}(y) = P(Y_{t_1k_1} \leq y)$ are the empirical cumulative distribution functions for the control and treatment group outcomes at time $t_1$ and outcome $k_1$. These covariances capture within-group dependencies, reflecting correlations across time and outcomes. They are critical for constructing the variance-covariance matrix $\bm{\Sigma}$ and determining the properties of the rank difference vector. The sample size ratio $\lambda = \frac{n_x}{n_y}$ denotes the control ($n_x$) to treatment ($n_y$) group ratio.

Rejection of the null hypothesis $H_0$ occurs at a significance level $\alpha$ if $T_{LRST} > z_\alpha$, where $z_\alpha$ is the $100(1-\alpha)\%$ quantile of a standard normal distribution. These variance components, $c_{t_1k_1t_2k_2}$ and $d_{t_1k_1t_2k_2}$, are essential for constructing the test statistic and estimating its power under the alternative hypothesis.

\subsection{Power and sample size calculation\label{subsec:22}}

\subsubsection{Estimators of variance components}
We begin by presenting the estimators of $c_{t_1k_1t_2k_2}$ and $d_{t_1k_1t_2k_2}$, as proposed in Xu \textit{et al}. \cite{xu2025SBR}: 

\begin{align*}
\hat{c}_{t_1k_1t_2k_2} &= \frac{1}{n_x} \sum_{i=1}^{n_x} \left\{ \left[\frac{1}{n_y} \sum_{j=1}^{n_y} I(y_{jt_1k_1} < x_{it_1k_1}) - \frac{1 - \hat{\theta}_{t_1k_1}}{2}\right] 
      \left[\frac{1}{n_y} \sum_{j=1}^{n_y} I(y_{jt_2k_2} < x_{it_2k_2}) - \frac{1 - \hat{\theta}_{t_2k_2}}{2}\right]  \right\}, \\
\hat{d}_{t_1k_1t_2k_2} &= \frac{1}{n_y} \sum_{j=1}^{n_y} \left\{ \left[\frac{1}{n_x} \sum_{i=1}^{n_x} I(x_{it_1k_1} < y_{jt_1k_1}) - \frac{1 + \hat{\theta}_{t_1k_1}}{2}\right] 
    \left[\frac{1}{n_x} \sum_{i=1}^{n_x} I(x_{it_2k_2} < y_{jt_2k_2}) - \frac{1 + \hat{\theta}_{t_2k_2}}{2}\right]  \right\},
\end{align*}
where $\hat{\theta}_{tk} = \frac{2}{N} \left(\Bar{R}_{y \cdot t k} - \Bar{R}_{x \cdot t k} \right)$. 

\subsubsection{Asymptotic properties of estimators}
To derive the asymptotic distributions of $\hat{c}_{t_1k_1t_2k_2}$ and $\hat{d}_{t_1k_1t_2k_2}$, along with their covariance structure, we define the following notations ($\mu_4(Z)$ is the fourth moment of a random variable $Z$):  
\begin{align*}
    \mathcal{K}_1^{(t_1,k_1)} &= \mu_4\left[G_{t_1k_1}(X_{t_1k_1})  \right]  - Var\left[G_{t_1k_1}(X_{t_1k_1}) \right]^2, \\
    \mathcal{K}_2^{(t_1,k_1)} &= \mu_4\left[F_{t_1k_1}(Y_{t_1k_1})  \right]  - Var\left[F_{t_1k_1}(Y_{t_1k_1}) \right]^2, \\
    \mathcal{L}_1^{(t_1,k_1,t_2,k_2)} &= Var\left[G_{t_1k_1}(X_{t_1k_1})\right]Var\left[G_{t_2k_2}(X_{t_2k_2})\right] + c_{t_1k_1t_2k_2}^2, \\
    \mathcal{L}_2^{(t_1,k_1,t_2,k_2)} &= Var\left[F_{t_1k_1}(Y_{t_1k_1})\right]Var\left[F_{t_2k_2}(Y_{t_2k_2})\right] + d_{t_1k_1t_2k_2}^2.
\end{align*}
Using these definitions, the following theorem provides the asymptotic distributions of these estimators:

\begin{theorem}
\label{thmcdhat}
Let \( \hat{c}_{t_1k_1t_2k_2} \) and \( \hat{d}_{t_1k_1t_2k_2} \) be estimators of the covariance components for the control and treatment groups, respectively. The asymptotic distributions of these estimators are as follows:

\[
    \sqrt{n_x}\left(\hat{c}_{t_1k_1t_2k_2} - c_{t_1k_1t_2k_2} \right) \xrightarrow{d}
\begin{cases}
    \mathcal{N}\left(0, \mathcal{K}_1^{(t_1,k_1)}\right), & \text{if } t_1 = t_2, \, k_1 = k_2 \\
    \mathcal{N}\left(0, \mathcal{L}_1^{(t_1,k_1,t_2,k_2)}\right), & \text{if } t_1 \neq t_2 \text{ or } k_1 \neq k_2
\end{cases}
\]

\[
    \sqrt{n_y}\left(\hat{d}_{t_1k_1t_2k_2} - d_{t_1k_1t_2k_2} \right) \xrightarrow{d}
\begin{cases}
    \mathcal{N}\left(0, \mathcal{K}_2^{(t_1,k_1)} \right), & \text{if } t_1 = t_2, \, k_1 = k_2 \\
    \mathcal{N}\left(0, \mathcal{L}_2^{(t_1,k_1,t_2,k_2)} \right), & \text{if } t_1 \neq t_2 \text{ or } k_1 \neq k_2
\end{cases}
\]

Here, \( \xrightarrow{d} \) denotes convergence in distribution, meaning that as the sample sizes \( n_x \) and \( n_y \) increase, the distributions of \( \hat{c}_{t_1k_1t_2k_2} \) and \( \hat{d}_{t_1k_1t_2k_2} \) approach the specified normal distributions.

Furthermore, for any \( t_1, k_1, t_2, k_2 \), the estimators \( \sqrt{N}\hat{c}_{t_1k_1t_2k_2} \) and \( \sqrt{N}\hat{d}_{t_1k_1t_2k_2} \) are asymptotically independent. The joint asymptotic distribution is given by:

\[
\begin{cases}
    \mathcal{N} \left( \sqrt{N} 
    \begin{pmatrix}
        c_{t_1k_1t_2k_2} \\
        d_{t_1k_1t_2k_2}
    \end{pmatrix},
    \begin{pmatrix}
        \left(1 + \frac{1}{\lambda}\right) \mathcal{K}_1^{(t_1,k_1)} & 0 \\
        0 & \left(1 + \lambda \right)\mathcal{K}_2^{(t_1,k_1)}
    \end{pmatrix}
    \right), & \text{if } t_1 = t_2, \, k_1 = k_2 \\
    
    \mathcal{N} \left( \sqrt{N}
    \begin{pmatrix}
        c_{t_1k_1t_2k_2} \\
        d_{t_1k_1t_2k_2}
    \end{pmatrix},
    \begin{pmatrix}
        \left(1 + \frac{1}{\lambda}\right) \mathcal{L}_1^{(t_1,k_1,t_2,k_2)} & 0 \\
        0 & \left(1 + \lambda\right)\mathcal{L}_2^{(t_1,k_1,t_2,k_2)}
    \end{pmatrix}
    \right), & \text{if } t_1 \neq t_2 \text{ or } k_1 \neq k_2
\end{cases}
\]
\qed
\end{theorem}

Detailed derivations and proofs for Theorem~\ref{thmcdhat} are provided in Appendix Section~\ref{Sec:app:Them1}.

Theorem~\ref{thmcdhat} provides the asymptotic distributions of the estimators $\hat{c}_{t_1k_1t_2k_2}$ and $\hat{d}_{t_1k_1t_2k_2}$. Based on this result, the estimate of $\Sigma_{t_1t_2}$ is given as:
\begin{align*}
\widehat{\Sigma}_{t_1t_2} = \frac{1}{K^2} \sum_{k_1,k_2=1}^K \left[\left( 1 + \frac{1}{\lambda}\right)\hat{c}_{t_1k_1t_2k_2} + (1+ \lambda)\hat{d}_{t_1k_1t_2k_2} \right], \quad \text{and} \quad \widehat{\bm{\Sigma}}_{T \times T} = \frac{\left(1+\lambda \right)\bm{J}^T (\hat{\bm{C}}+\lambda \hat{\bm{D}})\bm{J}}{\lambda}, 
\end{align*}
where $\bm{J}$ is a $T \times 1$ vector of $1$'s, and  $\hat{\bm{C}}$ and $\hat{\bm{D}}$ are estimators of the $T \times T$ matrices $\bm{C}$ and $\bm{D}$. The elements of $\bm{C}$ and $\bm{D}$ are
\begin{align*}
C_{t_1,t_2} = \frac{1}{K^2} \sum_{k_1=1}^K \sum_{k_2=1}^K c_{t_1k_1t_2k_2}, \quad D_{t_1,t_2} = \frac{1}{K^2}\sum_{k_1=1}^K \sum_{k_2=1}^K d_{t_1k_1t_2k_2}.
\end{align*}
From Theorem~\ref{thmcdhat}, and using the asymptotic joint density of $\hat{c}_{t_1k_1t_2k_2}$ 
and $\hat{d}_{t_1k_1t_2k_2}$, we can provide the asymptotic distribution of $\widehat{\Sigma}_{t_1k_1t_2k_2}$, as given in the following corollary.

\begin{corollary}
\label{corro1}
    By applying Theorem~\ref{thmcdhat}, we obtain the asymptotic distribution of
    \[
    \hat{\Sigma}_{t_1,t_2} = \frac{1}{K^2} \sum_{k_1=1}^K \sum_{k_2=1}^K \left[\left(1 + \frac{1}{\lambda}\right)\hat{c}_{t_1k_1t_2k_2} + (1 + \lambda)\hat{d}_{t_1k_1t_2k_2}\right].
    \]
    This distribution is asymptotically normal with mean $\Sigma_{t_1,t_2}$ and variance $Var_{\infty}(\widehat{\bm{\Sigma}})/N$. The exact expression for the variance is provided in Appendix Section~\ref{Sec:app:cor1}. Furthermore, the elements of $\sqrt{N}\hat{\bm{\Sigma}}$ are asymptotically independent. \qed
\end{corollary}
Detailed derivations and proofs for Corollary~\ref{corro1} are provided in Appendix Section~\ref{Sec:app:cor1}. From Corollary~\ref{corro1}, we see that $\widehat{\Sigma}_{t_1k_1t_2k_2}$ converges to a normal distribution with mean $\Sigma_{t_1k_1t_2k_2}$ and its asymptotic variance goes to $0$ as $N \rightarrow \infty$. Using Bienaymé-Chebyshev Inequality, we conclude that $\widehat{\Sigma}_{t_1k_1t_2k_2}$ converges to $\Sigma_{t_1k_1t_2k_2}$ in probability. 

\subsubsection{Power calculation}
Based on the asymptotic properties of $\widehat{\Sigma}$, we obtain the asymptotic distribution of $T_{LRST}$ using results from Xu \textit{et al.} \cite{xu2025SBR} and Slutsky's inequality:

\begin{align}
    T_{LRST} = \frac{\bar{R}_{y\cdot\cdot\cdot} - \bar{R}_{x\cdot\cdot\cdot}}{\sqrt{\widehat{var}(\bar{R}_{y\cdot\cdot\cdot} - \bar{R}_{x\cdot\cdot\cdot})}} \Rightarrow \mathcal{N}\left( \frac{T\sqrt{N}\bar{\theta}}{2\sqrt{\bm{J}^T \bm{\Sigma} \bm{J}}}, 1 \right),
    \label{eq:tasymp}
\end{align}
where $\bm{\Sigma} = \left(1+\lambda \right)\bm{J}^T (\hat{\bm{C}}+\lambda \hat{\bm{D}})\bm{J}/\lambda$. Under the null hypothesis ($\bar{\theta} = 0$), the distribution of $T_{LRST}$ is asymptotically standard normal. Under the alternative hypothesis ($\bar{\theta} \neq 0$), the test statistic $T_{LRST}$ follows a non-central normal distribution with a non-zero mean. Using Eq~\eqref{eq:tasymp}, we derive the theoretical power of the LRST test, as presented in Theorem~\ref{thmPower}. Additionally, the corresponding minimum sample size required to achieve the desired power is provided in Theorem~\ref{thm:sample}.

\begin{theorem} \label{thmPower}
The test statistic $T_{LRST}$ asymptotically follows a normal distribution with mean $\frac{\sqrt{N} T \bar{\theta}}{\sqrt{4T^2 \bm{J}^T \bm{\Sigma} \bm{J}}}$ and variance 1. The theoretical power of $T_{LRST}$, assuming $\bar{\theta} > 0$, is given by:
\begin{equation} \label{eq:theoPower}
\mathcal{P} = \Phi \left( \frac{\bar{\theta}}{\sqrt{\frac{4(1+\lambda) \bm{J}^T \left(\bm{C} + \lambda \bm{D}\right) \bm{J}}{N \lambda T^2}}} - z_\alpha \right),
\end{equation}
where $\Phi(\cdot)$ denotes the cumulative distribution function of the standard normal distribution, and $z_\alpha$ is the critical value at significance level $\alpha$. \qed
\end{theorem}

Detailed derivations and proofs for Theorem~\ref{thmPower} are provided in Appendix Section~\ref{Sec:app:Them2}. The expression for the theoretical power $\mathcal{P}$ depends on $\bar{\theta}$, $\bm{C}$, and $\bm{D}$, which in turn rely on the marginal distributions $G_{tk}$ and $F_{tk}$ of $X_{tk}$ and $Y_{tk}$, respectively. For instance, assuming normality of the marginal distributions, $\theta_{tk}$ can be computed as $P(X_{tk} < Y_{tk}) - P(X_{tk} > Y_{tk})$. Numerical integration or Monte Carlo techniques can be employed to compute $\theta_{tk}$, $c_{t_1k_1t_2k_2}$, and $d_{t_1k_1t_2k_2}$, which are then used to derive $\bar{\theta}$, $\bm{C}$, and $\bm{D}$ for theoretical power calculations under normality assumptions. Similar methods apply for other known pre-specified marginal distributions.

However, in practical applications, the true marginal distributions are rarely known in advance. Relying solely on normality assumptions may lead to inaccurate power estimations, particularly if the data exhibit significant deviations from linearity or normality. To address this, it is crucial to estimate power directly from the data, ensuring that the estimates are asymptotically consistent with the true power. The asymptotic normality of the estimated power is established in Theorem~\ref{thm4}, while Theorem~\ref{thm:sample} provides the framework for determining the minimum sample size required to achieve desired power levels. Together, these results offer a robust methodology for practical power and sample size calculations in real-world scenarios.

\subsubsection{Sample size calculation}
To ensure the LRST test achieves a specified power level ($\pi$), the minimum sample size required can be derived as follows:
\begin{theorem} \label{thm:sample}
The minimum sample size \( N \) required to achieve a power of at least \( \pi \) (assuming \( \bar{\theta} > 0 \)) is given by:
\begin{equation}
    N = 4 \frac{(1 + \lambda)}{\lambda T^2} \left[\frac{\Phi^{-1}(\pi) + z_{\alpha}}{\bar{\theta}} \right]^2 \left[ \bm{J}^T (\bm{C} + \lambda \bm{D}) \bm{J} \right],
    \label{eq:sampSize}
\end{equation}
where \( \Phi^{-1}(\pi) \) is the inverse cumulative distribution function of the standard normal distribution corresponding to the desired power \( \pi \), \( z_\alpha \) is the critical value for a significance level of \( \alpha \), \( \bar{\theta} \) is the overall treatment effect, \( T \) is the number of post-baseline time points, and \( \bm{C} \) and \( \bm{D} \) are the covariance matrices for the control and treatment groups, respectively.
\end{theorem}
Detailed derivations and proofs for Theorem~\ref{thm:sample} are provided in Appendix Section~\ref{Sec:app:Them3}. 

\subsection{Practical estimation of power for LRST} \label{subsec:23}

The theoretical power presented in Theorem~\ref{thmPower} depends on the unknown variance components, represented by the $\bm{C}$ and $\bm{D}$ matrices, which are not available in practical scenarios. Consequently, estimating power using real-world data is necessary. The power estimate is obtained by substituting empirical estimators of $\bar{\theta}$, $\bm{C}$, and $\bm{D}$ into Eq.~\eqref{eq:theoPower}. Using Theorems~\ref{thmPower} and~\ref{thmcdhat}, along with Corollary~\ref{corro1}, the asymptotic distribution of the power estimate can be derived. Specifically, the variance components $c_{t_1k_1t_2k_2}$ and $d_{t_1k_1t_2k_2}$ are estimated by their empirical counterparts, $\hat{c}_{t_1k_1t_2k_2}$ and $\hat{d}_{t_1k_1t_2k_2}$, respectively.

The overall treatment effect, $\bar{\theta} = \frac{1}{TK} \sum_{t=1}^T \sum_{k=1}^K \theta_{tk}$, is estimated as: $
\hat{\theta}_{tk} = \frac{1}{n_x n_y} \sum_{i=1}^{n_x} \sum_{j=1}^{n_y} \left[ I(x_{itk} < y_{jtk}) - I(x_{itk} > y_{jtk}) \right]$. By applying Lyapunov’s Central Limit Theorem, the asymptotic distribution of $\hat{\bar{\theta}}$ is obtained. The Delta method is then used to derive the asymptotic distribution of the estimated power, leading to the following result:

\begin{theorem} \label{thm4}
The estimated power of the LRST is given by: $\hat{\mathcal{P}} = \Phi \left( \frac{\hat{\bar{\theta}}}{\sqrt{\frac{4 \bm{J}^\top \hat{\bm{\Sigma}} \bm{J}}{N T^2}}} - z_{\alpha} \right)$, where $\Phi(\cdot)$ is the cumulative distribution function of the standard normal distribution, $\hat{\bar{\theta}}$ is the estimated overall treatment effect, $\hat{\bm{\Sigma}}$ is the estimated covariance matrix, and $z_{\alpha}$ is the critical value corresponding to a significance level $\alpha$.

The asymptotic distribution of $\hat{\mathcal{P}}$ follows: $\hat{\mathcal{P}} \sim \mathcal{N}(\mathcal{P}, \text{Var}(\hat{\mathcal{P}}))$, where the variance of $ \hat{\mathcal{P}}$ is:
\begin{align*}
\text{Var}\left(\hat{\mathcal{P}}\right) = N \phi^2\left(\frac{\bar{\theta}}{\sqrt{\frac{4 \bm{J}^\top \bm{\Sigma} \bm{J}}{N T^2}}} - z_{\alpha}\right) \frac{\bar{\theta}^2 T^2}{16 (\bm{J}^\top \bm{\Sigma} \bm{J})^3} \left[\frac{1}{K^4} \sum_{\substack{t_1 \neq t_2 \\ t_1, t_2 = 1}}^T \sum_{k_1, k_2 = 1}^K \mathcal{G}_{t_1k_1t_2k_2} + \frac{1}{K^4} \sum_{t=1}^T \left( \sum_{\substack{k_1 \neq k_2 \\ k_1, k_2 = 1}}^K \mathcal{G}_{tk_1tk_2} + \sum_{k=1}^K \mathcal{H}_{tktk} \right) \right],
\end{align*}
where $\phi(\cdot)$ is the standard normal density function. The variance components are captured by $\mathcal{G}_{t_1k_1t_2k_2}$ and $\mathcal{H}_{tktk}$, defined as:
\[
\mathcal{G}_{t_1k_1t_2k_2} = \left(1 + \frac{1}{\lambda}\right) \mathcal{L}_1^{(t_1, k_1, t_2, k_2)} + (1 + \lambda) \mathcal{L}_2^{(t_1, k_1, t_2, k_2)}, \quad
\mathcal{H}_{tktk} = \left(1 + \frac{1}{\lambda}\right) \mathcal{K}_1^{(t, k)} + (1 + \lambda) \mathcal{K}_2^{(t, k)}.
\]

Here, $\mathcal{G}_{t_1k_1t_2k_2}$ captures variance across different time points and outcomes, while $\mathcal{H}_{tktk}$ reflects within-outcome variance components.

\qed
\end{theorem}

Theorem~\ref{thm4} provides a practical framework for estimating the power of the LRST using real-world data. The variance components $\mathcal{G}_{t_1k_1t_2k_2}$ and $\mathcal{H}_{t_1k_1t_1k_1}$ account for correlations across outcomes and time points, ensuring robust power estimation. Detailed derivations and proofs for Theorem~\ref{thm4} are provided in Appendix Section~\ref{Sec:app:Them4}.

\section{Simulation Study\label{sec:sim}}
This simulation study evaluates the performance and theoretical properties of the LRST under conditions reflective of real-world clinical trials, with objectives including validating theoretical power calculations, comparing empirical and theoretical results, determining sample size requirements, and assessing the consistency of variance component estimators. Section~\ref{Sec:sim-setup} outlines the simulation setup, detailing the statistical model, parameters, and assumptions based on the Bapi 302 trial protocol. Section~\ref{Sec:sim-power} compares theoretical and empirical power, highlighting their convergence with increasing sample sizes, while Section~\ref{Sec:sim-SampleSize} explores sample size requirements for achieving specified power levels under different group size ratios. Finally, Section~\ref{Sec:sim-Variance} evaluates the accuracy and consistency of variance component estimators, providing further validation of the LRST framework.

\subsection{Simulation setup} \label{Sec:sim-setup}
The simulation study focuses on two longitudinal outcomes ($K=2$) measured over seven post-baseline visits ($T=7$). These outcomes represent typical clinical endpoints in RCTs. These outcomes are designed to reflect distinct mean trajectories for the placebo and treatment groups, modeling differential progression over time. The setup ensures the simulation mimics real-world scenarios, providing a robust test of the LRST framework. For the placebo group, the mean outcome trajectories across visits are:
\[
\bm{\mu}^{(1)} = \begin{pmatrix}
    0 & -1.38507 & -2.77014 & -4.15521 & -5.54028 & -6.92535 & -8.31042 \\
  0 & -2.65461 & -5.30922 & -7.96383 & -10.61844 & -13.27305 & -15.92766
\end{pmatrix}.
\]
For the treatment group, the progression is more gradual, reflecting the anticipated therapeutic effect:
\[
\bm{\mu}^{(2)} = \begin{pmatrix}
     0 & -1.016737 & -2.033473 & -3.05021 & -4.066947 & -5.083683 & -6.10042 \\
  0 & -1.757943 & -3.515887 & -5.27383 & -7.031773 & -8.789717 & -10.54766
\end{pmatrix}.
\]
Both groups exhibit increasing standard deviations over time, capturing the heteroscedasticity commonly observed in clinical trials:
\[
\bm{\sigma}^{(1)} = \bm{\sigma}^{(2)} = \begin{pmatrix}
    0 & 4.79 & 5.43 & 6.54 & 7.37 & 8.15 & 9.11 \\
    0 & 10.27 & 12.85 & 14.95 & 15.35 & 16.87 & 18.19
\end{pmatrix}
\]

The marginal distributions of the outcomes for the placebo and treatment groups are modeled as $F_{tk} = \mathcal{N}\left(\mu_{tk}^{(1)}, \sigma_{tk}^{(1)2} \right)$ and $G_{tk} = \mathcal{N}\left(\mu_{tk}^{(2)}, \sigma_{tk}^{(2)2} \right)$, respectively. A correlation of 0.5 between the two outcomes is included to account for dependencies between different clinical measures within subjects. The simulation parameters are derived from the BAPI 302 trial protocol, ensuring realism and relevance to clinical applications. This setup provides a rigorous framework to evaluate the performance of the LRST across varying sample sizes, outcome distributions, and levels of correlation, aiming to assess how well the test captures treatment effects and validates theoretical properties under practical conditions.

\subsection{Comparison of Theoretical, Empirical, and Estimated power} \label{Sec:sim-power}

To evaluate the accuracy of the theoretical power derived in Theorem~\ref{thmPower}, we compared it with empirical power from simulations and estimated power based on parameter estimation. Table~\ref{tab:powerComparison} summarizes the results across three metrics: Empirical Power, Theoretical Power, and Estimated Power. Empirical Power was obtained by applying the LRST to simulated datasets, while Theoretical Power was calculated using the closed-form expression from Theorem~\ref{thmPower}, assuming Gaussian marginal distributions that reflect the true data-generating process. This measure assumes known variance components and treatment effects. Estimated Power was computed by substituting estimated variance components and treatment effects into the theoretical power formula, as described in Theorem~\ref{thm4}. Standard errors for Estimated Power, based on multiple simulation replicates, are reported in parentheses.

The results in Table~\ref{tab:powerComparison} demonstrate strong agreement among the three measures, particularly as sample size increases. At $N = 500$, both Empirical and Theoretical Power converge to approximately 0.86–0.87, validating the theoretical framework. Estimated Power also closely aligns with these values, with minor deviations attributable to variability in parameter estimation. For smaller sample sizes, slight discrepancies emerge. For instance, at $N = 100$, the Estimated Power (0.42) slightly exceeds both the Empirical and Theoretical Power (0.35), likely due to greater uncertainty in estimating variance components and treatment effects, as reflected by the larger standard errors. As sample size increases, these discrepancies diminish, consistent with the asymptotic properties of the LRST. By $N = 300$, all three measures are nearly identical, with minimal variability. These findings highlight the practical utility of the theoretical power formula as a quick, reliable method for power calculation, particularly in the design phase of clinical trials. The Estimated Power approach adds flexibility by allowing researchers to leverage observed data when theoretical parameters are unavailable or derived from prior studies. The convergence of the three power measures with increasing sample sizes underscores the robustness and scalability of the LRST framework, making it a valuable tool for planning and interpreting clinical trials with multiple longitudinal outcomes.

\begin{table}[h!]
\centering
\begin{tabular}{cccc}
\hline
Sample Size ($N$) & Empirical Power & Theoretical Power & Estimated Power \\ \hline
100 & 0.35 & 0.35 & 0.42 (0.08) \\
300 & 0.68 & 0.70 & 0.68 (0.07) \\
500 & 0.86 & 0.87 & 0.85 (0.04) \\ \hline
\end{tabular}
\caption{Comparison of empirical, theoretical, and estimated power across sample sizes. Standard errors for the estimated power are shown in parentheses.}
\label{tab:powerComparison}
\end{table}

\subsection{Sample size determination for target power levels}\label{Sec:sim-SampleSize}

This subsection investigates the minimum sample sizes required to achieve specific power levels for detecting an effect size of (2.21, 5.38), as outlined in the Bapi 302 trial protocol. Table~\ref{tab:sampSize} presents the required sample sizes under two group size ratios: $\lambda = 2/3$ (smaller control group) and $\lambda = 1$ (equal group sizes). For lower power thresholds (e.g., 0.1), the required sample sizes are minimal, with only seven participants needed for both $\lambda = 2/3$ and $\lambda = 1$. However, as the desired power increases, the required sample size grows substantially. Achieving 80\% power requires 318 participants for $\lambda = 2/3$ and 305 participants for $\lambda = 1$, while 90\% power requires 443 and 423 participants, respectively. While the differences between these ratios are modest, they illustrate the efficiency gains associated with balanced group sizes. Slight imbalances, such as $\lambda = 2/3$, have minimal impact on total sample size, but more extreme imbalances would likely demand larger overall samples to maintain sufficient power.

These findings provide practical guidance for clinical trial planning. Using the formula from Theorem~\ref{thm:sample}, researchers can efficiently estimate minimum sample sizes required to achieve specific power levels, facilitating resource allocation while ensuring robust statistical power to detect treatment effects. By applying these results, trial designers can optimize cost, feasibility, and statistical rigor in longitudinal studies.

\begin{table}[h!]
    \centering
    \begin{tabular}{lccccccccc} \hline
        \textbf{Power} & \textbf{0.1} & \textbf{0.2} & \textbf{0.3} & \textbf{0.4} & \textbf{0.5} & \textbf{0.6} & \textbf{0.7} & \textbf{0.8} & \textbf{0.9} \\ \hline
        $\lambda = 2/3$ & 7 & 34 & 65 & 99 & 139 & 185 & 242 & 318 & 443 \\ 
        $\lambda = 1$ & 7 & 32 & 62 & 96 & 134 & 179 & 232 & 305 & 423 \\ \hline
    \end{tabular}
    \caption{Minimum sample sizes required to achieve target power levels for detecting an effect size of (2.21, 5.38), based on the Bapi 302 trial protocol. Results are provided for two group size ratios: $\lambda = 2/3$ (smaller control group) and $\lambda = 1$ (equal group sizes).}
    \label{tab:sampSize}
\end{table}

\subsection{Variance component validation} \label{Sec:sim-Variance}

This subsection evaluates the accuracy and convergence of the variance component estimators $\hat{C}_{t_1,t_2}$ and $\hat{D}_{t_1,t_2}$, as derived in Theorem~\ref{thmcdhat}. The performance of these estimators was assessed using two key metrics: mean squared error (MSE) and mean absolute error (MAE), defined as $\text{MSE} = \frac{1}{T^2} \sum_{t_1,t_2=1}^T \xi_{t_1,t_2}^2$ and $\text{MAE} = \frac{1}{T^2} \sum_{t_1,t_2=1}^T |\xi_{t_1,t_2}|$, where $\xi_{t_1,t_2} = \hat{C}_{t_1,t_2} - C_{t_1,t_2}$ or $\xi_{t_1,t_2} = \hat{D}_{t_1,t_2} - D_{t_1,t_2}$. These measures quantify the closeness of the estimated matrices $\hat{\bm{C}}$ and $\hat{\bm{D}}$ to their true counterparts $\bm{C}$ and $\bm{D}$. Table~\ref{tab:CD_error} presents MSE and MAE values across sample sizes $N = 100$, $300$, and $500$. The results demonstrate strong convergence properties of the estimators. For $\hat{\bm{C}}$, the MSE decreases from 0.0027 at $N = 100$ to 0.0016 at $N = 500$, while the MAE reduces from 0.041 to 0.032. Similarly, for $\hat{\bm{D}}$, the MSE declines from 0.002 to 0.001, and the MAE drops from 0.038 to 0.032 over the same sample sizes. These trends confirm that as the sample size increases, both estimators become more accurate, converging towards the true variance components.

\begin{table}[h!]
    \centering
    \begin{tabular}{ccc|ccc}
        \hline
        \multicolumn{3}{c|}{\textbf{$\hat{\bm{C}}$ (Variance Component)}} & \multicolumn{3}{c}{\textbf{$\hat{\bm{D}}$ (Variance Component)}} \\ \hline
        $N$ & MSE & MAE & $N$ & MSE & MAE \\ \hline
        100 & 0.0027 & 0.041 & 100 & 0.002 & 0.038 \\ 
        300 & 0.0018 & 0.034 & 300 & 0.001 & 0.033 \\ 
        500 & 0.0016 & 0.032 & 500 & 0.001 & 0.032 \\ \hline
    \end{tabular}
    \caption{Mean squared error (MSE) and mean absolute error (MAE) for the variance component estimators $\hat{\bm{C}}$ and $\hat{\bm{D}}$ across sample sizes $N = 100$, $300$, and $500$. Both MSE and MAE decrease with increasing sample size, validating the accuracy and convergence of the estimators.}
    \label{tab:CD_error}
\end{table}

To further validate the asymptotic properties of these estimators, we compared the estimated asymptotic standard errors of $\hat{C}_{t_1,t_2}$ and $\hat{D}_{t_1,t_2}$ with their theoretical counterparts, as derived in Theorem~\ref{thmcdhat}. Table~\ref{tab:CD_error_full} reports the results for $N = 500$, with each matrix entry displaying the estimated standard error alongside the theoretical standard error in parentheses. The results show excellent agreement between estimated and theoretical values, confirming the reliability of the asymptotic approximations. For example, at $(t_1, t_2) = (1, 1)$, the estimated standard error for $\hat{C}_{t_1,t_2}$ is 0.019, matching the theoretical value exactly. Similarly, the estimated standard error for $\hat{D}_{t_1,t_2}$ at $(t_1, t_2) = (1, 1)$ is 0.018, closely aligning with the theoretical value of 0.020. This pattern of close alignment persists across all matrix entries, underscoring the consistency and robustness of the variance estimators.

The results from Tables~\ref{tab:CD_error} and \ref{tab:CD_error_full} provide compelling evidence for the validity of the LRST framework in estimating variance components for multiple longitudinal outcomes. The decreasing MSE and MAE values, combined with the close agreement between estimated and theoretical standard errors, affirm the reliability of the methodology. These findings highlight the practical utility of the LRST in clinical trials and longitudinal studies, ensuring precise variance estimation to support robust statistical inference.

\begin{table}[h!]
    \centering
    \begin{tabular}{c}
        \begin{tabular}{cccccc}
            \hline
            \multicolumn{6}{c}{\textbf{Standard Error of $\hat{C}_{t_1,t_2}$ for $N=500$ (Theoretical Standard Errors in Parentheses)}} \\ \hline
            0.019 (0.019) & 0.016 (0.016) & 0.016 (0.016) & 0.015 (0.015) & 0.015 (0.013) & 0.014 (0.011) \\ 
            0.016 (0.016) & 0.019 (0.019) & 0.015 (0.015) & 0.016 (0.015) & 0.015 (0.013) & 0.015 (0.012) \\ 
            0.016 (0.016) & 0.016 (0.016) & 0.018 (0.019) & 0.015 (0.014) & 0.015 (0.013) & 0.015 (0.012) \\ 
            0.015 (0.015) & 0.016 (0.015) & 0.015 (0.014) & 0.017 (0.019) & 0.015 (0.012) & 0.015 (0.010) \\ 
            0.015 (0.014) & 0.015 (0.013) & 0.015 (0.013) & 0.015 (0.012) & 0.016 (0.019) & 0.014 (0.009) \\ 
            0.014 (0.012) & 0.015 (0.012) & 0.015 (0.011) & 0.015 (0.011) & 0.014 (0.009) & 0.016 (0.018) \\ \hline
        \end{tabular}
        \\
        \begin{tabular}{cccccc}
            \hline
            \multicolumn{6}{c}{\textbf{Standard Error of $\hat{D}_{t_1,t_2}$ for $N=500$ (Theoretical Standard Errors in Parentheses)}} \\ \hline
            0.018 (0.020) & 0.014 (0.017) & 0.014 (0.016) & 0.013 (0.015) & 0.013 (0.014) & 0.013 (0.012) \\ 
            0.014 (0.017) & 0.018 (0.020) & 0.014 (0.016) & 0.014 (0.015) & 0.014 (0.013) & 0.013 (0.012) \\ 
            0.014 (0.016) & 0.014 (0.016) & 0.016 (0.020) & 0.013 (0.014) & 0.013 (0.013) & 0.013 (0.011) \\ 
            0.013 (0.015) & 0.014 (0.015) & 0.013 (0.014) & 0.016 (0.020) & 0.013 (0.012) & 0.013 (0.011) \\ 
            0.013 (0.014) & 0.014 (0.013) & 0.013 (0.013) & 0.013 (0.012) & 0.016 (0.020) & 0.014 (0.009) \\ 
            0.013 (0.012) & 0.013 (0.012) & 0.013 (0.011) & 0.013 (0.011) & 0.014 (0.009) & 0.015 (0.019) \\ \hline
        \end{tabular}
    \end{tabular}
    \caption{Comparison of estimated and theoretical asymptotic standard errors for $\hat{\bm{C}}_{t_1,t_2}$ and $\hat{\bm{D}}_{t_1,t_2}$ at $N=500$. The close alignment between estimated and theoretical values confirms the consistency and accuracy of the variance component estimators.}
    \label{tab:CD_error_full}
\end{table}

\section{Real Data Analysis\label{sec:realData}}
This section applies the LRST methodology to real-world clinical trial data to evaluate its practical utility. Two notable studies, one focused on AD and the other on PD, were analyzed to assess the LRST's performance in detecting treatment effects across multiple longitudinal outcomes. These analyses highlight the strengths of the LRST framework and provide insights into its application for complex neurodegenerative conditions.

\subsection{Analysis of Bapineuzumab (Bapi) 302 trial}

\begin{figure}[h!]
    \centering
    \includegraphics[width=0.8\linewidth]{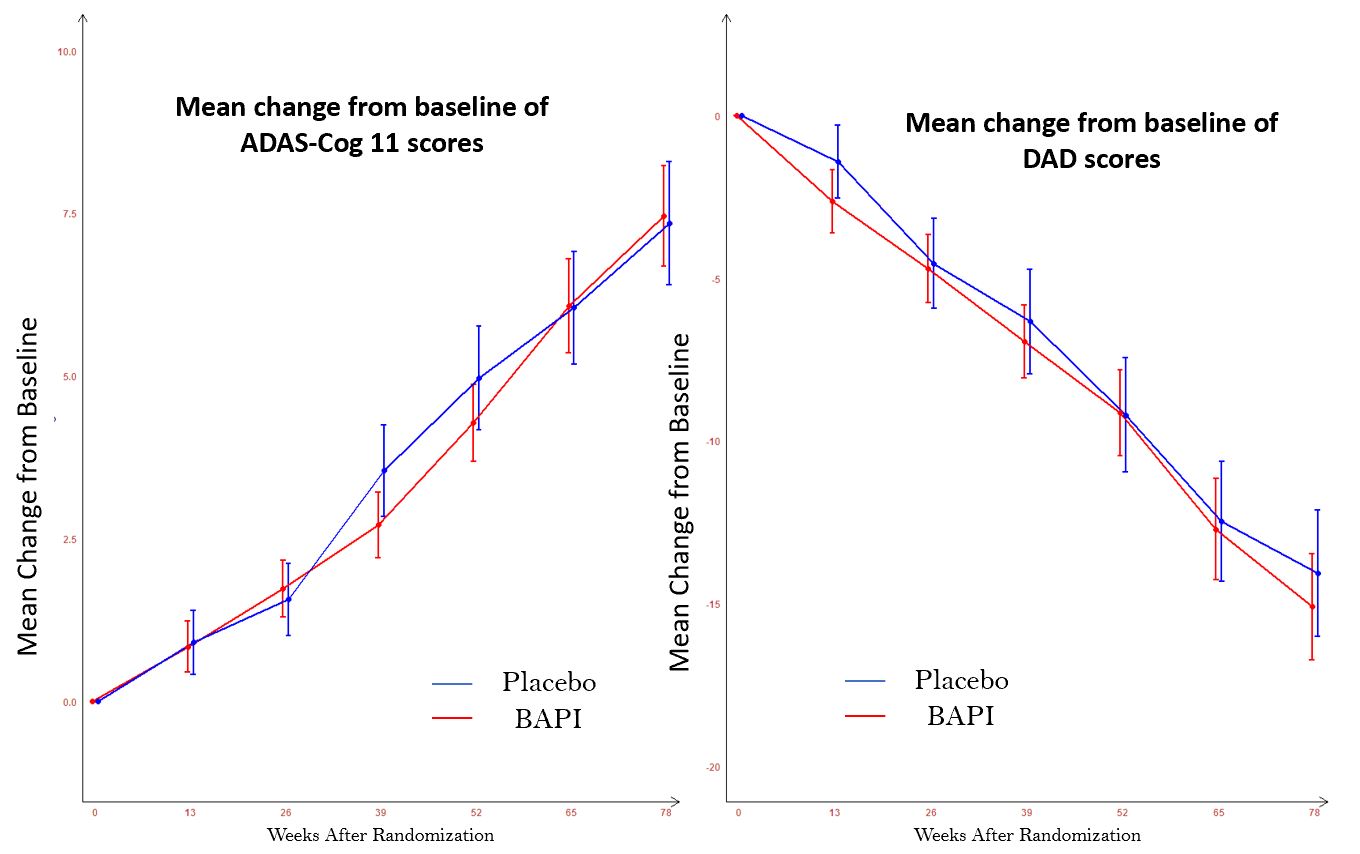}
    \caption{Mean changes from baseline to week 78 in ADAS-cog11 (left) and DAD (right) scores in the Bapineuzumab 302 Trial. Higher ADAS-cog11 scores indicate greater cognitive impairment, while higher DAD scores reflect better functional independence. Vertical lines represent 95\% confidence intervals.}
    \label{fig:mean-curve}
\end{figure}

The Bapineuzumab (Bapi) 302 trial \cite{salloway2014two} was a Phase 3, multicenter, double-blind, randomized controlled trial designed to evaluate the efficacy of intravenous Bapi compared to placebo in patients with mild-to-moderate AD. Conducted across 170 sites in the United States from December 2007 to April 2012, the trial randomized 759 participants in a 2:3 ratio into placebo ($n_x = 311$) and treatment ($n_y = 448$) groups. Assessments were performed every 13 weeks over a 78-week period ($t = 13, 26, 39, 52, 65, 78$, $T = 6$). The study evaluated two primary outcomes: the Alzheimer’s Disease Assessment Scale-Cognitive Subscale (ADAS-cog11) and the Disability Assessment for Dementia (DAD). ADAS-cog11 consists of 11 tasks assessing cognitive domains, including memory, language, and praxis, with scores ranging from 0 to 70; higher scores indicate greater cognitive impairment. The DAD scale evaluates functional independence based on caregiver-reported activities of daily living, such as grooming and meal preparation, with scores ranging from 0 to 100, where higher scores reflect better functional ability. The mean changes in ADAS-cog11 and DAD scores over the 78 weeks are illustrated in Figure~\ref{fig:mean-curve}. Both groups showed consistent cognitive and functional decline, evidenced by increasing ADAS-cog11 scores (worsening cognition) and decreasing DAD scores (reduced daily functioning). No significant differences were observed between the treatment and placebo groups across either outcome.

The LRST was applied to assess the global treatment effect of Bapineuzumab. The analysis yielded a test statistic of $-0.0474$ (SE: $1.950$), corresponding to a Z-statistic of $-0.024$ ($P = 0.509$), indicating no significant overall treatment efficacy. To further evaluate the power of the test, the global treatment effect parameter, $\bar{\theta}$, was estimated using both a theoretical approach, assuming normality of the marginal distributions, and an empirical approach based on observed data. Theoretical calculations yielded $\bar{\theta} = -0.02$, while the empirical estimate was $\hat{\bar{\theta}} = -0.07$. The negative values suggest a treatment effect opposite to the expected benefit, indicating that Bapineuzumab did not improve cognitive or functional outcomes. Both theoretical power (Theorem~\ref{thmPower}) and estimated power (Theorem~\ref{thm4}) were low, at 0.030 and 0.028, respectively. These results provide no statistical evidence of an overall treatment benefit for Bapineuzumab in patients with mild-to-moderate AD.

\subsection{Analysis of the Azilect study}

The Azilect study \citep{Hattori2019PRD} was a Phase 3, randomized, double-blind, placebo-controlled trial evaluating the efficacy of rasagiline in Japanese patients with early-stage PD. Participants aged 30 to 79, diagnosed within the previous five years, were randomized 1:1 to receive either 1 mg/day of rasagiline or placebo over a 26-week treatment period. Clinical assessments were conducted at baseline and weeks 6, 10, 14, 20, and 26 ($T = 5$). The primary endpoint was the Movement Disorder Society Unified Parkinson’s Disease Rating Scale (MDS-UPDRS), a widely used tool for assessing Parkinsonian symptoms in clinical trials \citep{Regnault2019JN}. The MDS-UPDRS includes 65 items across four domains: Part I (Non-Motor Aspects of Daily Living, 13 items), Part II (Motor Aspects of Daily Living, 13 items), Part III (Motor Examination, 33 items), and Part IV (Motor Complications, 6 items). Each item is scored on a 5-point Likert scale (0–4), with higher scores indicating greater symptom severity \citep{Regnault2019JN, Goetz2008MDS}.

\begin{figure}[h!]
    \centering
    \includegraphics[width=0.8\linewidth]{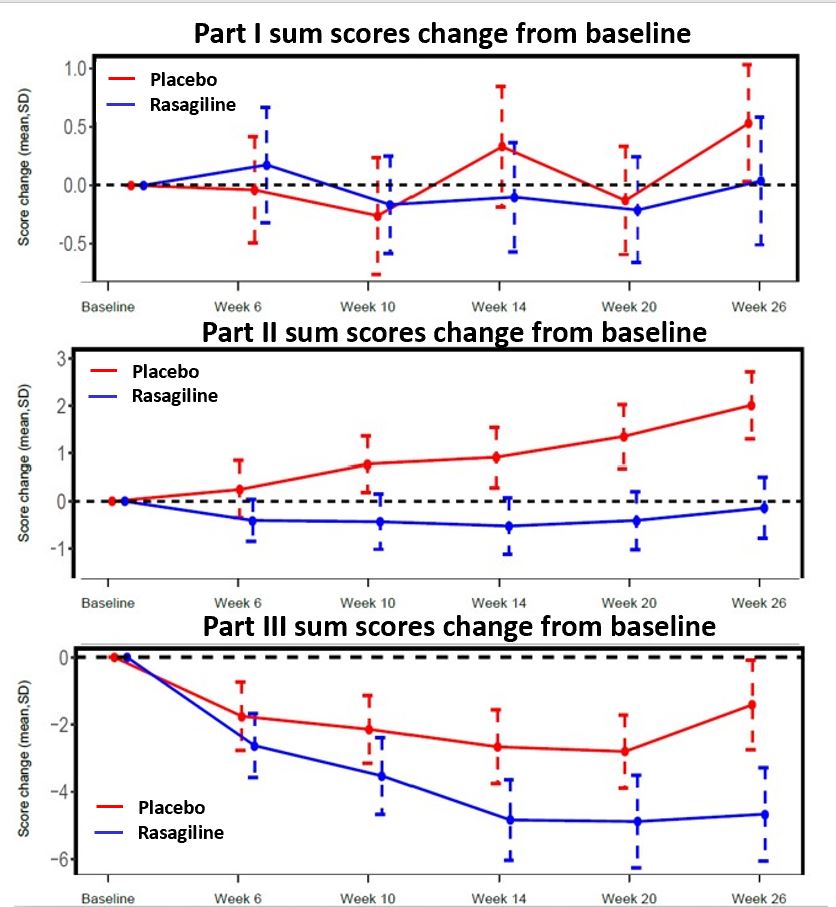}
    \caption{Mean changes from baseline to each visit in MDS-UPDRS Parts I, II, and III total scores. The placebo group (red squares) and rasagiline group (blue circles) are plotted with means and standard deviations. Vertical lines represent standard deviations, and the horizontal dashed line denotes the baseline. Negative score changes indicate symptom improvement. The top panel shows Part I (Non-Motor Aspects of Daily Living), the middle panel shows Part II (Motor Aspects of Daily Living), and the bottom panel shows Part III (Motor Examination).}
    \label{score}
\end{figure}

Figure~\ref{score} illustrates distinct trends across the three MDS-UPDRS components. In Part I (top panel), non-motor symptoms fluctuate around baseline in both groups with no clear separation. In Part II (middle panel), the placebo group exhibits increasing symptom severity, while the rasagiline group shows initial improvement that diminishes toward the trial's end. In Part III (bottom panel), both groups show sustained motor function improvement, with the rasagiline group demonstrating a consistently larger benefit. To standardize interpretation, total scores were adjusted so that higher values correspond to better outcomes by reversing the sign of the ordinal item scores (multiplied by $-1$). The LRST yielded a test statistic of $4.929$ (SE: $1.572$), corresponding to a Z-statistic of $3.135$ ($P = 8.592 \times 10^{-4}$), indicating a significant global treatment effect of rasagiline across the three outcomes.

The effect size $\hat{\bar{\theta}}$ was estimated under two approaches: assuming normality of the marginal distributions and using the mean rank difference from the data. The effect size was $0.234$ under normality and $0.948$ when estimated empirically. Variance components $\bm{C}$ and $\bm{D}$ were derived using both theoretical (under normality) and empirical (from data) methods. The $\bm{C}$ and $\hat{\bm{C}}$ matrices are presented below, with differences attributed to small sample size and deviations from normality. Theoretical power under the Gaussian assumption was 0.39, while the estimated power using empirical data was 0.99, indicating substantial deviation from normality. To achieve 90\% power under normality, 676 participants were required, but the observed sample size of 209 achieved 99\% power. Using empirical estimates, only 154 participants were needed for 90\% power, suggesting the trial was sufficiently powered even with a smaller sample size. These results confirm a significant treatment effect of rasagiline on motor symptoms in early-stage PD. The discrepancy in required sample sizes underscores the impact of distributional assumptions on power calculations, highlighting the value of nonparametric approaches such as the LRST in longitudinal clinical trials.

\begin{align*}
    C = \begin{pmatrix}
        0.32 & 0.27 & 0.23 & 0.22 & 0.20 \\
        0.27 & 0.23 & 0.20 & 0.18 & 0.17 \\
        0.23 & 0.20 & 0.17 & 0.16 & 0.15 \\
        0.22 & 0.18 & 0.16 & 0.15 & 0.14 \\
        0.20 & 0.17 & 0.15 & 0.14 & 0.13
    \end{pmatrix},
    \quad
    \hat{C} = \begin{pmatrix}
        0.28 & 0.10 & 0.14 & 0.12 & 0.12 \\
        0.19 & 0.30 & 0.21 & 0.18 & 0.16 \\
        0.14 & 0.21 & 0.31 & 0.19 & 0.20 \\
        0.12 & 0.18 & 0.19 & 0.26 & 0.20 \\
        0.12 & 0.16 & 0.20 & 0.20 & 0.32
    \end{pmatrix}
\end{align*}

\section{Conclusion}\label{dec:conclusion}

This article develops a robust framework for power and sample size calculations tailored to the LRST, addressing a critical gap in nonparametric methodologies for analyzing multivariate longitudinal data. By integrating theoretical derivations, asymptotic properties, and practical estimation techniques, we provide a comprehensive approach for designing clinical trials that assess complex temporal outcomes in neurodegenerative diseases. Our results show that the LRST framework reliably estimates power and determines appropriate sample sizes across diverse clinical trial settings. Validation through numerical simulations and real-world applications, including the Bapineuzumab and Azilect trials, demonstrates its practical utility and reliability. The analysis of real data highlights the LRST’s strengths in detecting treatment effects across multiple longitudinal outcomes while also revealing limitations in scenarios with small effect sizes or deviations from standard distributional assumptions. These findings emphasize the importance of incorporating both theoretical models and empirical insights when applying the LRST to clinical data.

This work advances nonparametric inference by offering a systematic approach to evaluating treatment efficacy that accounts for temporal dynamics and multivariate structures. Future research could focus on extending the LRST framework to address missing data, a frequent and complex challenge in longitudinal studies. Techniques such as multiple imputation or inverse probability weighting could be integrated into the LRST to handle incomplete data while preserving the nonparametric advantages of the method. Additionally, adapting the LRST for high-dimensional outcomes, where the number of measurements exceeds the sample size, would enhance its applicability in modern clinical trials that involve comprehensive biomarker panels, imaging data, or multi-omics datasets. This could involve incorporating regularization techniques or dimensionality reduction methods to maintain statistical efficiency. Refining variance estimation techniques to improve robustness under varying data conditions, such as non-normal distributions or heteroscedasticity, would further strengthen the LRST's performance. Moreover, the development of adaptive trial designs that allow for interim analyses and sample size recalibration based on accumulating data could optimize resource use while ensuring adequate statistical power. Finally, integrating machine learning approaches with the LRST framework might provide novel insights into complex treatment effects and facilitate more personalized medicine applications.

In conclusion, the proposed framework marks a significant advancement in nonparametric methods for multivariate longitudinal data analysis. It lays a foundation for future methodological innovations and practical applications in the design and analysis of clinical trials targeting complex, progressive diseases, ensuring that robust statistical tools keep pace with the evolving demands of modern biomedical research.

\section*{Acknowledgements}
This study, carried out under YODA Project \#2020-4448, used data of the Bapineuzumab 302 study obtained from the Yale University Open Data Access Project, which has an agreement with JANSSEN RESEARCH \& DEVELOPMENT, L.L.C.. The interpretation and reporting of research using this data are solely the responsibility of the authors and do not necessarily represent the official views of the Yale University Open Data Access Project or JANSSEN RESEARCH \& DEVELOPMENT, L.L.C..

The Azilect study data was accessed from the Critical Path Institute’s Critical Path for Parkinson’s Consortium. The Critical Path Institute’s CPP Consortium is funded by Parkinson’s United Kingdom and the following industry members: AbbVie, Biogen, MSD, Takeda, Sanofi, Roche, IXICO, Clario, Novartis, AnnovisBio, Neuron23, VanquaBio, Bayer and UCB. We also acknowledge additional CPP member organizations, including the Parkinson’s Disease Foundation, The Michael J. Fox Foundation, the Davis Phinney Foundation, The Cure Parkinson’s Trust, PMD Alliance, the University of Oxford, University of Cambridge, Newcastle University, University of Glasgow, as well as the NINDS, US Food and Drug Administration, and the European Medicines Agency.

\section*{Conflict of interest}
The authors declare no potential conflict of interest.

\section*{Supporting Information}
None.


\bibliography{ref, finalRef}

\appendix


\section*{Appendix A: Proof of Theorem~\ref{thmcdhat}}  
\label{Sec:app:Them1}

\subsection*{A.1 Introduction and Setup}

The goal of Theorem~\ref{thmcdhat} is to establish the asymptotic distributions of the covariance component estimators $ \hat{c}_{t_1k_1t_2k_2} $ and $ \hat{d}_{t_1k_1t_2k_2} $. These estimators quantify the covariance structures in the Longitudinal Rank Sum Test (LRST) for multivariate longitudinal data.

We aim to show that:
\[
\sqrt{n_x}\left(\hat{c}_{t_1k_1t_2k_2} - c_{t_1k_1t_2k_2}\right) \xrightarrow{d} 
\begin{cases}
    \mathcal{N}\left(0, \mathcal{K}_1^{(t_1,k_1)}\right), & \text{if } t_1 = t_2, k_1 = k_2 \\
    \mathcal{N}\left(0, \mathcal{L}_1^{(t_1,k_1,t_2,k_2)}\right), & \text{otherwise}
\end{cases}
\]
and similarly for $ \hat{d}_{t_1k_1t_2k_2} $. Here, the notation $ \xrightarrow{d} $ denotes convergence in distribution.

We begin by defining the estimators and simplifying the notation to facilitate the derivation of their asymptotic properties.

\subsection*{A.2 Estimation and Asymptotic Distribution of $c_{t_1k_1t_2k_2}$}

We aim to estimate the covariance component $c_{t_1k_1t_2k_2} = \text{Cov}\left(G_{t_1k_1}(X_{t_1k_1}), G_{t_2k_2}(X_{t_2k_2})\right)$ using the following estimator:
\[
\hat{c}_{t_1k_1t_2k_2} = \frac{1}{n_x} \sum_{u=1}^{n_x} \left[\frac{1}{n_y} \sum_{v=1}^{n_y} I(y_{vt_1k_1} < x_{ut_1k_1}) - \frac{1 - \hat{\theta}_{t_1k_1}}{2} \right] 
\left[\frac{1}{n_y} \sum_{v=1}^{n_y} I(y_{vt_2k_2} < x_{ut_2k_2}) - \frac{1 - \hat{\theta}_{t_2k_2}}{2} \right],
\]
where $I(\cdot)$ is the indicator function, and $\hat{\theta}_{tk}$ represents the estimated treatment effect at time $t$ and outcome $k$.

Using the relationship 
$\frac{1 - \hat{\theta}_{tk}}{2} = \frac{1}{n_x n_y} \sum_{i=1}^{n_x} \sum_{j=1}^{n_y} I(y_{jtk} < x_{itk})$, we can rewrite $\hat{c}_{t_1k_1t_2k_2}$ as:
\[
\hat{c}_{t_1k_1t_2k_2} = \frac{1}{n_x} \sum_{u=1}^{n_x} 
\left[\frac{1}{n_y} \sum_{v=1}^{n_y} I(y_{vt_1k_1} < x_{ut_1k_1}) - E\left(\frac{1}{n_y} \sum_{v=1}^{n_y} I(y_{vt_1k_1} < x_{ut_1k_1})\right)\right]
\]
\[
\quad \times 
\left[\frac{1}{n_y} \sum_{v=1}^{n_y} I(y_{vt_2k_2} < x_{ut_2k_2}) - E\left(\frac{1}{n_y} \sum_{v=1}^{n_y} I(y_{vt_2k_2} < x_{ut_2k_2})\right)\right] + o(1),
\]
where $o(1)$ denotes a term that converges to zero as the sample size increases.

To simplify notation, we define:
$\mathcal{U}_{tk} = \frac{1}{n_y} \sum_{v=1}^{n_y} I(y_{vtk} < x_{utk}), \quad \tilde{\mathcal{U}}_{tk} = \mathcal{U}_{tk} - E(\mathcal{U}_{tk})$. Thus, the estimator simplifies to:
$\hat{c}_{t_1k_1t_2k_2} \approx \frac{1}{n_x} \sum_{u=1}^{n_x} \tilde{\mathcal{U}}_{t_1k_1} \tilde{\mathcal{U}}_{t_2k_2}$.

\textbf{Asymptotic Distribution:}  

When $t_1 \neq t_2$ or $k_1 \neq k_2$, applying the well-known asymptotic results for sample covariance, we derive that $\hat{c}_{t_1k_1t_2k_2}$ is asymptotically normal with mean $c_{t_1k_1t_2k_2}$ and variance: 
$\frac{\sigma_{\mathcal{U}_{t_1k_1}}^2 \sigma_{\mathcal{U}_{t_2k_2}}^2 + \sigma_{\mathcal{U}_{t_1k_1}\mathcal{U}_{t_2k_2}}^2}{n_x}$,
where:
\[
\sigma_{\mathcal{U}_{t_1k_1}}^2 = \text{Var}\left[G_{t_1k_1}(X_{t_1k_1})\right], \quad \sigma_{\mathcal{U}_{t_2k_2}}^2 = \text{Var}\left[G_{t_2k_2}(X_{t_2k_2})\right],
\]
and
\[
\sigma_{\mathcal{U}_{t_1k_1}\mathcal{U}_{t_2k_2}} = \text{Cov}\left[G_{t_1k_1}(X_{t_1k_1}), G_{t_2k_2}(X_{t_2k_2})\right] = c_{t_1k_1t_2k_2}.
\]
Hence, for $t_1 \neq t_2$ or $k_1 \neq k_2$, the asymptotic distribution of $\hat{c}_{t_1k_1t_2k_2}$ can be expressed as:
\[
\hat{c}_{t_1k_1t_2k_2} \xrightarrow{d} \mathcal{N}\left(c_{t_1k_1t_2k_2}, \frac{\text{Var}\left[G_{t_1k_1}(X_{t_1k_1})\right]\text{Var}\left[G_{t_2k_2}(X_{t_2k_2})\right] + c_{t_1k_1t_2k_2}^2}{n_x}\right).
\]

When $t_1 = t_2$ and $k_1 = k_2$, the asymptotic distribution follows from the properties of sample variance. In this case, $\hat{c}_{t_1k_1t_2k_2}$ is asymptotically normal with mean $c_{t_1k_1t_2k_2}$ and variance:
\[
\mu_4\left[G_{t_1k_1}(X_{t_1k_1})\right] - \left(\text{Var}\left[G_{t_1k_1}(X_{t_1k_1})\right]\right)^2,
\]
where $\mu_4(\cdot)$ denotes the fourth central moment.

Thus, the full asymptotic distribution of $\hat{c}_{t_1k_1t_2k_2}$ is:
\[
\sqrt{n_x}\left(\hat{c}_{t_1k_1t_2k_2} - c_{t_1k_1t_2k_2}\right) \xrightarrow{d}
\begin{dcases}
    \mathcal{N}\left(0, \mu_4\left[G_{t_1k_1}(X_{t_1k_1})\right] - \left(\text{Var}\left[G_{t_1k_1}(X_{t_1k_1})\right]\right)^2 \right), & \text{if } t_1 = t_2, k_1 = k_2 \\
    \mathcal{N}\left(0, \text{Var}\left[G_{t_1k_1}(X_{t_1k_1})\right] \text{Var}\left[G_{t_2k_2}(X_{t_2k_2})\right] + c_{t_1k_1t_2k_2}^2 \right), & \text{otherwise}.
\end{dcases}
\]

Here, $\xrightarrow{d}$ denotes convergence in distribution, implying that as the sample size increases, the distribution of $\sqrt{n_x}\left(\hat{c}_{t_1k_1t_2k_2} - c_{t_1k_1t_2k_2}\right)$ approaches a normal distribution.

\subsection*{A.3 Estimation and Asymptotic Distribution of $d_{t_1k_1t_2k_2}$}

The estimation and asymptotic derivation for the covariance component $d_{t_1k_1t_2k_2} = \text{Cov}\left(F_{t_1k_1}(Y_{t_1k_1}), F_{t_2k_2}(Y_{t_2k_2})\right)$ follows a process analogous to that outlined for $c_{t_1k_1t_2k_2}$ in Section A.2. The estimator is given by:
\[
\hat{d}_{t_1k_1t_2k_2} = \frac{1}{n_y} \sum_{v=1}^{n_y} \left[\frac{1}{n_x} \sum_{u=1}^{n_x} I(x_{ut_1k_1} < y_{vt_1k_1}) - \frac{1 - \hat{\theta}_{t_1k_1}}{2} \right]
\left[\frac{1}{n_x} \sum_{u=1}^{n_x} I(x_{ut_2k_2} < y_{vt_2k_2}) - \frac{1 - \hat{\theta}_{t_2k_2}}{2} \right].
\]

By applying similar asymptotic principles and the Central Limit Theorem for sample covariances, the asymptotic distribution of $\hat{d}_{t_1k_1t_2k_2}$ is:

\[
\sqrt{n_y}\left(\hat{d}_{t_1k_1t_2k_2} - d_{t_1k_1t_2k_2}\right) \xrightarrow{d}
\begin{dcases}
    \mathcal{N}\left(0, \mu_4\left[F_{t_1k_1}(Y_{t_1k_1})\right] - \left(\text{Var}\left[F_{t_1k_1}(Y_{t_1k_1})\right]\right)^2 \right), & \text{if } t_1 = t_2, k_1 = k_2, \\
    \mathcal{N}\left(0, \text{Var}\left[F_{t_1k_1}(Y_{t_1k_1})\right] \text{Var}\left[F_{t_2k_2}(Y_{t_2k_2})\right] + d_{t_1k_1t_2k_2}^2 \right), & \text{otherwise}.
\end{dcases}
\]

The notation and convergence results are consistent with those provided in Section A.2 for $c_{t_1k_1t_2k_2}$. The primary difference lies in the substitution of the marginal distributions of $X_{tk}$ with those of $Y_{tk}$.

\subsection*{A.4 Joint Asymptotic Distribution of $\hat{c}_{t_1k_1t_2k_2}$ and $\hat{d}_{t_1k_1t_2k_2}$}

Next, we will derive the covariance between $\hat{c}_{t_1k_1t_2k_2}$ and $\hat{d}_{t_1k_1t_2k_2}$. First let us define $\mathcal{V}_{tk}$ analogous to $\mathcal{U}_{tk}$: $\mathcal{V}_{tk} = \frac{1}{n_y} \sum_{v=1}^{n_y} I(x_{vtk} < y_{utk})
$.

\begin{align}
    \nonumber Cov\left(\hat{c}_{t_1k_1t_2k_2}, \hat{d}_{t_1k_1t_2k_2} \right)
    &= Cov\left\{\frac{1}{n_x}\sum_{u=1}^{n_x}\left[\mathcal{U}_{t_1k_1} - E \left(\mathcal{U}_{t_1k_1} \right)\right]\left[\mathcal{U}_{t_2k_2} - E \left(\mathcal{U}_{t_2k_2}\right) \right], \right. \\
    & \left. \quad \quad \quad \quad \frac{1}{n_y} \sum_{v=1}^{n_y} \left[\mathcal{V}_{t_1k_1} - E \left(\mathcal{V}_{t_1k_1} \right)\right]\left[\mathcal{V}_{t_2k_2} - E \left(\mathcal{V}_{t_2k_2}\right) \right]\right\} \\
    &= \nonumber \frac{1}{n_xn_y} \sum_{u=1}^{n_x}\sum_{v=1}^{n_y}E \Bigg\{\left[\mathcal{U}_{t_1k_1} - E \left(\mathcal{U}_{t_1k_1} \right)\right]\left[\mathcal{U}_{t_2k_2} - E \left(\mathcal{U}_{t_2k_2}\right) \right] 
    \\
    & \quad \quad \quad 
    \left[\mathcal{V}_{t_1k_1} - E \left(\mathcal{V}_{t_1k_1} \right)\right]\left[\mathcal{V}_{t_2k_2} - E \left(\mathcal{V}_{t_2k_2} \right)\right]\Bigg\}  -  c_{t_1k_1t_2k_2}d_{t_1k_1t_2k_2} \label{eq1} \\
    \nonumber &= \frac{1}{n_x n_y} \sum_{u=1}^{n_x}\sum_{v=1}^{n_y} E \left(\tilde{\mathcal{U}}_{t_1k_1} \tilde{\mathcal{U}}_{t_2k_2} \tilde{\mathcal{V}}_{t_1k_1} \tilde{\mathcal{V}}_{t_2k_2}\right) - c_{t_1k_1t_2k_2}d_{t_1k_1t_2k_2},
\end{align}
where $\tilde{\mathcal{U}}_{tk} =  \mathcal{U}_{tk} - E \left(\mathcal{U}_{tk}\right)$ and $\tilde{\mathcal{V}}_{tk} =  \mathcal{V}_{tk} - E \left(\mathcal{V}_{tk}\right)$.

To further analyze this, consider the interaction between the first and third elements:
\begin{align*}
    \tilde{\mathcal{U}}_{t_1k_1} \tilde{\mathcal{V}}_{t_1k_1} &=\left[\mathcal{U}_{t_1k_1} - E \left(\mathcal{U}_{t_1k_1}\right) \right]\left[\mathcal{V}_{t_1k_1} - E\left( \mathcal{V}_{t_1k_1}\right) \right] \\
     &= \left\{\frac{1}{n_y} \sum_{a=1}^{n_y} \left[I(y_{at_1k_1} < x_{ut_1k_1}) - E\left( \mathcal{U}_{t_1k_1}\right) \right] \right\}\left\{\frac{1}{n_x} \sum_{b=1}^{n_x} \left[I(y_{vt_1k_1} > x_{bt_1k_1}) - E \left(\mathcal{V}_{t_1k_1}\right) \right] \right\}
\end{align*}

Dependencies arise only when $a = v$ and $b = u$. To isolate these, rewrite the expression:

\begin{align*}
    &\left[\mathcal{U}_{t_1k_1} - E \left(\mathcal{U}_{t_1k_1}\right)\right] \left[\mathcal{U}_{t_2k_2} - E \left(\mathcal{U}_{t_2k_2}\right)\right] \\
    &= \left\{\frac{1}{n_y} \sum_{a=1, a \neq v}^{n_y} \left[I(y_{at_1k_1} < x_{ut_1k_1}) - E \left(\mathcal{U}_{t_1k_1}\right)\right] + \frac{1}{n_y} \left[I(y_{vt_1k_1} < x_{ut_1k_1}) - E \left(\mathcal{U}_{t_1k_1}\right)\right]\right\} \\
    &\quad \times \left\{\frac{1}{n_x} \sum_{b=1, b \neq u}^{n_x} \left[I(y_{vt_1k_1} > x_{bt_1k_1}) - E \left(\mathcal{V}_{t_1k_1}\right)\right] + \frac{1}{n_x} \left[I(y_{vt_1k_1} > x_{ut_1k_1}) - E \left(\mathcal{V}_{t_1k_1}\right)\right]\right\}.
\end{align*}

Let us introduce the following notations:

\begin{align*}
    \mathcal{A}_{tk} &= \frac{1}{n_y} \sum_{a=1, a \neq v}^{n_y} \left[ I(y_{atk} < x_{utk}) - E\left(\mathcal{U}_{tk}\right) \right], \quad  
    \mathcal{C}_{tk} = \frac{1}{n_y} \left[ I(y_{vtk} < x_{utk}) - E\left(\mathcal{U}_{tk}\right) \right], \\
    \mathcal{B}_{tk} &= \frac{1}{n_x} \sum_{b=1, b \neq u}^{n_x} \left[ I(y_{vtk} > x_{btk}) - E\left(\mathcal{V}_{tk}\right) \right], \quad  
    \mathcal{D}_{tk} = \frac{1}{n_x} \left[ I(y_{vtk} > x_{utk}) - E\left(\mathcal{V}_{tk}\right) \right].
\end{align*}

Using these notations, the expectation term in the covariance expression can be written as:

\begin{align*}
    E \left(\tilde{\mathcal{U}}_{t_1k_1} \tilde{\mathcal{U}}_{t_2k_2} \tilde{\mathcal{V}}_{t_1k_1} \tilde{\mathcal{V}}_{t_2k_2}\right) 
    &= E \left[ \left( \mathcal{A}_{t_1k_1} + \mathcal{C}_{t_1k_1} \right)
    \left( \mathcal{A}_{t_2k_2} + \mathcal{C}_{t_2k_2} \right)
    \left( \mathcal{B}_{t_1k_1} + \mathcal{D}_{t_1k_1} \right)
    \left( \mathcal{B}_{t_2k_2} + \mathcal{D}_{t_2k_2} \right) \right] \\
    &= E \left[ \left( \mathcal{A}_{t_1k_1} + \mathcal{C}_{t_1k_1} \right)
    \left( \mathcal{B}_{t_1k_1} + \mathcal{D}_{t_1k_1} \right)
    \left( \mathcal{A}_{t_2k_2} + \mathcal{C}_{t_2k_2} \right)
    \left( \mathcal{B}_{t_2k_2} + \mathcal{D}_{t_2k_2} \right) \right] \\
    &= E \left[ \left( \mathcal{A}_{t_1k_1} \mathcal{B}_{t_1k_1} 
    + \mathcal{A}_{t_1k_1} \mathcal{D}_{t_1k_1} 
    + \mathcal{B}_{t_1k_1} \mathcal{C}_{t_1k_1} 
    + \mathcal{C}_{t_1k_1} \mathcal{D}_{t_1k_1} \right) \right. \\
    & \quad \quad \left. \times \left( \mathcal{A}_{t_2k_2} \mathcal{B}_{t_2k_2} 
    + \mathcal{A}_{t_2k_2} \mathcal{D}_{t_2k_2} 
    + \mathcal{B}_{t_2k_2} \mathcal{C}_{t_2k_2} 
    + \mathcal{C}_{t_2k_2} \mathcal{D}_{t_2k_2} \right) \right].
\end{align*}

Since $E(\mathcal{A}_{t_1k_1}) = E(\mathcal{A}_{t_2k_2}) = E(\mathcal{B}_{t_1k_1}) = E(\mathcal{B}_{t_2k_2}) = 0$, and considering the independence between the components, we have:

\begin{align*}
    E\left( \mathcal{A}_{t_1k_1}\mathcal{B}_{t_1k_1}\mathcal{A}_{t_2k_2}\mathcal{D}_{t_2k_2}\right) 
    &= E\left( \mathcal{A}_{t_1k_1}\mathcal{B}_{t_1k_1}\mathcal{B}_{t_2k_2}\mathcal{C}_{t_2k_2}\right) 
    = E\left( \mathcal{A}_{t_1k_1}\mathcal{B}_{t_1k_1}\mathcal{C}_{t_2k_2}\mathcal{D}_{t_2k_2}\right) = 0, \\
    E\left( \mathcal{A}_{t_1k_1}\mathcal{D}_{t_1k_1}\mathcal{A}_{t_2k_2}\mathcal{B}_{t_2k_2}\right) 
    &= E\left( \mathcal{A}_{t_1k_1}\mathcal{D}_{t_1k_1}\mathcal{B}_{t_2k_2}\mathcal{C}_{t_2k_2}\right) 
    = E\left( \mathcal{A}_{t_1k_1}\mathcal{D}_{t_1k_1}\mathcal{C}_{t_2k_2}\mathcal{D}_{t_2k_2}\right) = 0, \\
    E\left( \mathcal{B}_{t_1k_1}\mathcal{C}_{t_1k_1}\mathcal{A}_{t_2k_2}\mathcal{B}_{t_2k_2}\right) 
    &= E\left( \mathcal{B}_{t_1k_1}\mathcal{C}_{t_1k_1}\mathcal{A}_{t_2k_2}\mathcal{D}_{t_2k_2}\right) 
    = E\left( \mathcal{B}_{t_1k_1}\mathcal{C}_{t_1k_1}\mathcal{C}_{t_2k_2}\mathcal{D}_{t_2k_2}\right) = 0, \\
    E\left( \mathcal{C}_{t_1k_1}\mathcal{D}_{t_1k_1}\mathcal{A}_{t_2k_2}\mathcal{B}_{t_2k_2}\right) 
    &= E\left( \mathcal{C}_{t_1k_1}\mathcal{D}_{t_1k_1}\mathcal{A}_{t_2k_2}\mathcal{D}_{t_2k_2}\right) 
    = E\left( \mathcal{C}_{t_1k_1}\mathcal{D}_{t_1k_1}\mathcal{B}_{t_2k_2}\mathcal{C}_{t_2k_2}\right) = 0.
\end{align*}

Therefore, the final remaining terms are:

\begin{align*}
    E\left(\tilde{\mathcal{U}}_{t_1k_1} \tilde{\mathcal{U}}_{t_2k_2} \tilde{\mathcal{V}}_{t_1k_1} \tilde{\mathcal{V}}_{t_2k_2}\right) 
    &= E\left( \mathcal{A}_{t_1k_1}\mathcal{B}_{t_1k_1}\mathcal{A}_{t_2k_2}\mathcal{B}_{t_2k_2} \right) 
    + E\left( \mathcal{A}_{t_1k_1}\mathcal{D}_{t_1k_1}\mathcal{A}_{t_2k_2}\mathcal{D}_{t_2k_2} \right) \\
    & \quad + E\left( \mathcal{B}_{t_1k_1}\mathcal{C}_{t_1k_1}\mathcal{B}_{t_2k_2}\mathcal{C}_{t_2k_2} \right) 
    + E\left( \mathcal{C}_{t_1k_1}\mathcal{D}_{t_1k_1}\mathcal{C}_{t_2k_2}\mathcal{D}_{t_2k_2} \right).
\end{align*}

Note that $E \left(\mathcal{A}_{t_1k_1}\mathcal{B}_{t_1k_1}\mathcal{A}_{t_2k_2}\mathcal{B}_{t_2k_2} \right)
= E\left( \mathcal{A}_{t_1k_1}\mathcal{A}_{t_2k_2}\right) E\left(\mathcal{B}_{t_1k_1}\mathcal{B}_{t_2k_2}\right) 
\rightarrow c_{t_1k_1t_2k_2} d_{t_1k_1t_2k_2}$, which cancels out the second term in Equation \ref{eq1}. Therefore, the only non-zero contributions to the covariance come from the remaining three terms, which we address individually. Let $F_{t_1k_1t_2k_2}$ and $G_{t_1k_1t_2k_2}$ denote the joint distributions of $(X_{t_1k_1}, X_{t_2k_2})$ and $(Y_{t_1k_1}, Y_{t_2k_2})$, respectively. We begin by evaluating $E \left(\mathcal{C}_{t_1k_1}\mathcal{C}_{t_2k_2}\right)$:

\begin{align*}
E \left(\mathcal{C}_{t_1k_1}\mathcal{C}_{t_2k_2}\right) 
&= \frac{1}{n_y^2} E\left[\left(I(y_{vt_1k_1} < x_{ut_1k_1}) - E\left(\mathcal{U}_{t_1k_1}\right)\right) 
\left(I(y_{vt_2k_2} < x_{ut_2k_2}) - E\left(\mathcal{U}_{t_2k_2}\right)\right)\right] \\
&= \frac{1}{n_y^2} \Bigg\{ E\left[I\left(y_{1t_1k_1}< x_{1t_1k_1}, y_{1t_2k_2}< x_{1t_2k_2}\right)\right] 
- E\left(\mathcal{U}_{t_2k_2}\right) E\left[I\left(y_{1t_1k_1}< x_{1t_1k_1}\right)\right] \\
&\quad - E\left(\mathcal{U}_{t_1k_1}\right) E\left[I\left(y_{1t_2k_2}< x_{1t_2k_2}\right)\right] 
+ E\left(\mathcal{U}_{t_1k_1}\right) E\left(\mathcal{U}_{t_2k_2}\right)\Bigg\} \\
&= \frac{1}{n_y^2} \Bigg\{ E\left[G_{t_1k_1t_2k_2}(X_{t_1k_1}, X_{t_2k_2})\right] 
- E\left[G_{t_1k_1}(X_{t_1k_1})\right] E\left[G_{t_2k_2}(X_{t_2k_2})\right]\Bigg\}.
\end{align*}

Similarly, we obtain:

\begin{align*}
E \left(\mathcal{D}_{t_1k_1}\mathcal{D}_{t_2k_2}\right) 
= \frac{1}{n_x^2} \Bigg\{ E\left[F_{t_1k_1t_2k_2}(Y_{t_1k_1}, Y_{t_2k_2})\right] 
- E\left[F_{t_1k_1}(Y_{t_1k_1})\right] E\left[F_{t_2k_2}(Y_{t_2k_2})\right]\Bigg\}.
\end{align*}

Therefore:

\begin{align*}
E \left(\mathcal{A}_{t_1k_1}\mathcal{A}_{t_2k_2} \mathcal{D}_{t_1k_1}\mathcal{D}_{t_2k_2}\right) 
&= E \left(\mathcal{A}_{t_1k_1}\mathcal{A}_{t_2k_2}\right) E\left(\mathcal{D}_{t_1k_1}\mathcal{D}_{t_2k_2}\right) \\
&\rightarrow c_{t_1k_1t_2k_2} \frac{1}{n_x^2} \Bigg\{ E\left[F_{t_1k_1t_2k_2}(Y_{t_1k_1}, Y_{t_2k_2})\right] 
- E\left[F_{t_1k_1}(Y_{t_1k_1})\right] E\left[F_{t_2k_2}(Y_{t_2k_2})\right]\Bigg\}, \\
E \left(\mathcal{B}_{t_1k_1}\mathcal{B}_{t_2k_2} \mathcal{C}_{t_1k_1}\mathcal{C}_{t_2k_2}\right) 
&= E \left(\mathcal{B}_{t_1k_1}\mathcal{B}_{t_2k_2}\right) E\left(\mathcal{C}_{t_1k_1}\mathcal{C}_{t_2k_2}\right) \\
&\rightarrow d_{t_1k_1t_2k_2} \frac{1}{n_y^2} \Bigg\{ E\left[G_{t_1k_1t_2k_2}(X_{t_1k_1}, X_{t_2k_2})\right] 
- E\left[G_{t_1k_1}(X_{t_1k_1})\right] E\left[G_{t_2k_2}(X_{t_2k_2})\right]\Bigg\}.
\end{align*}

The final term is:

\begin{align*}
E \left(\mathcal{C}_{t_1k_1}\mathcal{C}_{t_2k_2}\mathcal{D}_{t_1k_1}\mathcal{D}_{t_2k_2}\right) 
= \frac{1}{n_x^2 n_y^2} E\Bigg[\prod_{i=1}^2 \left( I(y_{vt_ik_i} < x_{ut_ik_i}) - E\left(\mathcal{U}_{t_ik_i}\right) \right) 
\left( I(y_{vt_ik_i} > x_{ut_ik_i}) - E\left(\mathcal{V}_{t_ik_i}\right) \right)\Bigg] 
= O\left(\frac{1}{n_x^2 n_y^2}\right).
\end{align*}

Thus, with $N = n_x + n_y$, the covariance becomes:

\begin{align*}
Cov\left(\sqrt{N} \hat{c}_{t_1k_1t_2k_2}, \sqrt{N} \hat{d}_{t_1k_1t_2k_2}\right) 
&= \frac{N}{n_y^2} \Bigg\{ E\left[G_{t_1k_1t_2k_2}(X_{t_1k_1}, X_{t_2k_2})\right] 
- E\left[G_{t_1k_1}(X_{t_1k_1})\right] E\left[G_{t_2k_2}(X_{t_2k_2})\right]\Bigg\} \\
&\quad + \frac{N}{n_x^2} \Bigg\{ E\left[F_{t_1k_1t_2k_2}(Y_{t_1k_1}, Y_{t_2k_2})\right] 
- E\left[F_{t_1k_1}(Y_{t_1k_1})\right] E\left[F_{t_2k_2}(Y_{t_2k_2})\right]\Bigg\} \\
&\quad + O\left(\frac{N}{n_x^2 n_y^2}\right) \\
&= O\left[\frac{1}{n_y}(1 + \lambda)\right] 
+ O\left[\frac{1}{n_x}(1 + 1/\lambda)\right] 
+ O\left(\frac{1}{n_x n_y^2} + \frac{1}{n_x^2 n_y}\right) \rightarrow 0.
\end{align*}

Hence, $\sqrt{N}\hat{c}_{t_1k_1t_2k_2}$ and $\sqrt{N}\hat{d}_{t_1k_1t_2k_2}$ are asymptotically independent, completing the proof of the theorem.

\section{Proof of Corollary~\ref{corro1}} 
\label{Sec:app:cor1}

Using techniques similar to those employed in the proof of Theorem~\ref{thmPower}, we can demonstrate that the summands of $\hat{\bm{\Sigma}}_{t_1,t_2}$, specifically $\hat{c}_{t_1k_1t_2k_2}$ and $\hat{d}_{t_1k_1t_2k_2}$, are independent of any other quadruple $(t_1k_3t_2k_4)$. As a result, the asymptotic distribution is derived by summing asymptotically independent normal distributions.

For $t_1 \neq t_2$, the asymptotic variance is given by:

\begin{align*}
  \frac{1}{K^4} \sum_{k_1=1}^K \sum_{k_2=1}^K &\left[\left(1 + \frac{1}{\lambda}\right) \left\{ \text{Var}\left[G_{t_1k_1}(X_{t_1k_1})\right] \text{Var}\left[G_{t_2k_2}(X_{t_2k_2})\right] + c_{t_1k_1t_2k_2}^2 \right\} \right. \\
  & \left. \quad + (1 + \lambda) \left\{ \text{Var}\left[F_{t_1k_1}(Y_{t_1k_1})\right] \text{Var}\left[F_{t_2k_2}(Y_{t_2k_2})\right] + d_{t_1k_1t_2k_2}^2 \right\} \right].
\end{align*}

For $t_1 = t_2$, the variance becomes:

\begin{align*}
    &\frac{1}{K^4} \sum_{\substack{k_1,k_2=1 \\ k_1 \neq k_2}}^K \Bigg[\left(1 + \frac{1}{\lambda}\right) \left\{ \text{Var}\left[G_{t_1k_1}(X_{t_1k_1})\right] \text{Var}\left[G_{t_1k_2}(X_{t_1k_2})\right] + c_{t_1k_1t_1k_2}^2 \right\} \\
    &\quad + (1 + \lambda) \left\{ \text{Var}\left[F_{t_1k_1}(Y_{t_1k_1})\right] \text{Var}\left[F_{t_1k_2}(Y_{t_1k_2})\right] + d_{t_1k_1t_1k_2}^2 \right\} \Bigg] \\
    &\quad + \frac{1}{K^4} \sum_{k_1=1}^K \left[\mu_4\left(G_{t_1k_1}(X_{t_1k_1})\right) - \text{Var}\left(G_{t_1k_1}(X_{t_1k_1})\right)^2 + \mu_4\left(F_{t_1k_1}(Y_{t_1k_1})\right) - \text{Var}\left(F_{t_1k_1}(Y_{t_1k_1})\right)^2 \right].
\end{align*}

Define the following quantities for simplicity:

\begin{align*}
  \mathcal{G}_{t_1k_1t_2k_2} &= \left(1 + \frac{1}{\lambda}\right) \left\{ \text{Var}\left[G_{t_1k_1}(X_{t_1k_1})\right] \text{Var}\left[G_{t_2k_2}(X_{t_2k_2})\right] + c_{t_1k_1t_2k_2}^2 \right\} \\
  &\quad + (1 + \lambda) \left\{ \text{Var}\left[F_{t_1k_1}(Y_{t_1k_1})\right] \text{Var}\left[F_{t_2k_2}(Y_{t_2k_2})\right] + d_{t_1k_1t_2k_2}^2 \right\}, \\
  \mathcal{H}_{t_1k_1} &= \mu_4\left[G_{t_1k_1}(X_{t_1k_1})\right] - \text{Var}\left[G_{t_1k_1}(X_{t_1k_1})\right]^2 + \mu_4\left[F_{t_1k_1}(Y_{t_1k_1})\right] - \text{Var}\left[F_{t_1k_1}(Y_{t_1k_1})\right]^2.
\end{align*}

The asymptotic variance of $\sqrt{N} \hat{\bm{\Sigma}}_{t_1t_2}$ is thus:

\[
\text{Var}_{\infty}\left(\hat{\bm{\Sigma}}_{t_1t_2}\right) = 
\begin{cases}
   \frac{1}{K^4} \sum_{k_1=1}^K \sum_{k_2=1}^K \mathcal{G}_{t_1k_1t_2k_2}, & \text{if } t_1 \neq t_2, \\
   \frac{1}{K^4} \left(\sum_{\substack{k_1, k_2=1 \\ k_1 \neq k_2}}^K \mathcal{G}_{t_1k_1t_1k_2} + \sum_{k_1=1}^K \mathcal{H}_{t_1k_1}\right), & \text{if } t_1 = t_2.
\end{cases}
\]

Since all elements are asymptotically independent, the distribution of $\bm{J}^\top \hat{\bm{\Sigma}} \bm{J}$ is asymptotically normal with mean $\bm{J}^\top \hat{\bm{\Sigma}} \bm{J}$ and variance given by:

\[
\text{Var}_{\infty}(\bm{J}^\top \hat{\bm{\Sigma}} \bm{J}) = \frac{1}{K^4} \sum_{\substack{t_1 \neq t_2 \\ t_1, t_2 = 1}}^T \sum_{k_1=1}^K \sum_{k_2=1}^K \mathcal{G}_{t_1k_1t_2k_2} 
+ \frac{1}{K^4} \sum_{t=1}^T \left( \sum_{\substack{k_1 \neq k_2 \\ k_1, k_2=1}}^K \mathcal{G}_{t k_1 t k_2} + \sum_{k=1}^K \mathcal{H}_{t k} \right).
\]

To prove the second part, we must demonstrate that the elements of $\hat{\bm{\Sigma}}_{t_1, t_2}$ are asymptotically independent. The elements of $\sqrt{N}\hat{\bm{\Sigma}}_{t_1, t_2}$ can be expressed as: 

\[
\hat{\bm{\Sigma}}_{t_1, t_2} = \frac{1}{K^2} \sum_{k_1=1}^K \sum_{k_2=1}^K \left( \lambda' \hat{c}_{t_1k_1t_2k_2} + \lambda \hat{d}_{t_1k_1t_2k_2} \right),
\]
where $\lambda' = 1 + \frac{1}{\lambda}$.

Expanding this expression gives:

\[
\hat{\bm{\Sigma}}_{t_1, t_2} = \frac{1}{K^2} \sum_{k_1=1}^K \sum_{k_2=1}^K \left\{ 
\lambda' \frac{1}{n_x} \sum_{u=1}^{n_x} \left[\mathcal{U}_{t_1k_1} - E \left(\mathcal{U}_{t_1k_1}\right)\right] \left[\mathcal{U}_{t_2k_2} - E \left(\mathcal{U}_{t_2k_2}\right)\right] 
+ \lambda \frac{1}{n_y} \sum_{v=1}^{n_y} \left[\mathcal{V}_{t_1k_1} - E \left(\mathcal{V}_{t_1k_1}\right)\right] \left[\mathcal{V}_{t_2k_2} - E \left(\mathcal{V}_{t_2k_2}\right)\right] 
\right\}.
\]

The covariance between two generic elements, $\hat{\bm{\Sigma}}_{t_1, t_2}$ and $\hat{\bm{\Sigma}}_{t_3, t_4}$, is then:

\[
\text{Cov} \left(\hat{\bm{\Sigma}}_{t_1, t_2}, \hat{\bm{\Sigma}}_{t_3, t_4} \right) = \frac{1}{K^4} \sum_{k_1=1}^K \sum_{k_2=1}^K \text{Cov} \left( \lambda' \hat{c}_{t_1k_1t_2k_2} + \lambda \hat{d}_{t_1k_1t_2k_2}, \lambda' \hat{c}_{t_3k_1t_4k_2} + \lambda \hat{d}_{t_3k_1t_4k_2} \right).
\]

As previously established, the asymptotic covariance of each pair $\hat{c}_{t_1k_1t_2k_2}$ and $\hat{d}_{t_1k_1t_2k_2}$ with any distinct quadruple $(t_3, k_1, t_4, k_2)$ vanishes due to asymptotic independence. Therefore, $\text{Cov} \left(\hat{\bm{\Sigma}}_{t_1, t_2}, \hat{\bm{\Sigma}}_{t_3, t_4} \right) = 0$.

This confirms that all elements of $\hat{\bm{\Sigma}}$ are asymptotically independent.

\section{Proof of Theorem~\ref{thmPower}} \label{Sec:app:Them2}

It has been established by Xue et. al. \cite{xu2025SBR} that the rank difference vector $ \frac{1}{\sqrt{N}} \bm{R} $ asymptotically follows a multivariate normal distribution with mean vector $ \frac{\sqrt{N}}{2} \bm{\theta} $ and covariance matrix $ \bm{\Sigma} $, as defined previously. Here, $ \bm{\theta} $ represents the vector of treatment effects at each time point and outcome, and $ \bm{\Sigma} $ captures the covariance structure of the rank differences.

The LRST statistic, defined in Equation~\eqref{eq:tlrst}, can be expressed as 
\[
T_{LRST} = \frac{\bar{R}_{y \cdot \cdot \cdot} - \bar{R}_{x \cdot \cdot \cdot}}{\sqrt{\widehat{\text{Var}}(\bar{R}_{y \cdot \cdot \cdot} - \bar{R}_{x \cdot \cdot \cdot})}} = \frac{\frac{1}{T} \bm{J}^\top \frac{\bm{R}}{\sqrt{N}}}{\sqrt{\widehat{\text{Var}}\left(\frac{1}{T} \bm{J}^\top \frac{\bm{R}}{\sqrt{N}}\right)}} = \frac{\bm{J}^\top \frac{\bm{R}}{\sqrt{N}}}{\sqrt{\bm{J}^\top \hat{\bm{\Sigma}} \bm{J}}},
\]
where $ \bm{J} $ is a contrast vector (with elements summing to zero), summarizing the rank differences across all time points and outcomes, and $ \hat{\bm{\Sigma}} $ is the estimated covariance matrix of $ \bm{R} $.

Applying the asymptotic result from Xu et. al.\cite{xu2025SBR} and Slutsky's theorem, we have 
\[
T_{LRST} \xrightarrow[]{d} \mathcal{N}\left(\frac{T \sqrt{N} \bar{\theta}}{2 \sqrt{\bm{J}^\top \bm{\Sigma} \bm{J}}}, 1\right),
\]
where $ \bar{\theta} = \frac{1}{T} \bm{J}^\top \bm{\theta} $ represents the average treatment effect over time and outcomes. This result indicates that, under the alternative hypothesis $ H_a $, the test statistic follows a normal distribution with a non-zero mean that depends on $ \bar{\theta} $, the sample size $ N $, and the covariance structure $ \bm{\Sigma} $.

The power of the test under $ H_a $ is then calculated as 
\[
\mathcal{P} = P_{H_a}(T_{LRST} > z_\alpha) = \Phi\left(\frac{\frac{T \sqrt{N} \bar{\theta}}{2 \sqrt{\bm{J}^\top \bm{\Sigma} \bm{J}}} - z_\alpha}{1}\right) = \Phi\left(\frac{T \sqrt{N} \bar{\theta}}{2 \sqrt{\bm{J}^\top \bm{\Sigma} \bm{J}}} - z_\alpha\right),
\]
where $ z_\alpha $ is the upper $ \alpha $-quantile of the standard normal distribution.

To further simplify, we substitute the structure of the covariance matrix $ \bm{\Sigma} = \frac{1 + \lambda}{\lambda} (\bm{C} + \lambda \bm{D}) $, where $ \bm{C} $ and $ \bm{D} $ capture within- and between-group variances, respectively, and $ \lambda $ represents the allocation ratio of the two groups. Therefore, 
\[
\bm{J}^\top \bm{\Sigma} \bm{J} = \frac{1 + \lambda}{\lambda} \bm{J}^\top (\bm{C} + \lambda \bm{D}) \bm{J}.
\]

Substituting this into the power expression gives 
\[
\mathcal{P} = \Phi\left(\frac{\bar{\theta}}{\sqrt{\frac{4(1 + \lambda) \bm{J}^\top (\bm{C} + \lambda \bm{D}) \bm{J}}{N \lambda T^2}}} - z_\alpha\right),
\]
which completes the proof. This final expression illustrates how the power depends on the treatment effect $ \bar{\theta} $, sample size $ N $, allocation ratio $ \lambda $, and the variance components $ \bm{C} $ and $ \bm{D} $.






\section{Proof of Theorem~\ref{thm:sample}} \label{Sec:app:Them3}

To ensure a minimum power of $ \pi $ (assuming $ \bar{\theta} > 0 $), the sample size $ N $ must satisfy $ \mathcal{P} \geq \pi $. 

Substituting the expression for power from Theorem~\ref{thmPower}, we have 
\[
\Phi \left( \frac{\bar{\theta}}{\sqrt{\frac{4(1+\lambda) \bm{J}^\top \left(\bm{C} + \lambda \bm{D}\right) \bm{J}}{N \lambda T^2}}} - z_\alpha \right) \geq \pi.
\]

Applying the inverse cumulative distribution function $ \Phi^{-1} $ to both sides gives 
\[
\frac{\bar{\theta}}{\sqrt{\frac{4(1+\lambda) \bm{J}^\top \left(\bm{C} + \lambda \bm{D}\right) \bm{J}}{N \lambda T^2}}} \geq z_\alpha + \Phi^{-1}(\pi).
\]

Squaring both sides and solving for $ N $ yields:
\begin{align*}
    \frac{4(1+\lambda) \bm{J}^\top \left(\bm{C} + \lambda \bm{D}\right) \bm{J}}{N \lambda T^2} &\leq \left( \frac{\bar{\theta}}{z_\alpha + \Phi^{-1}(\pi)} \right)^2, \\
    N &\geq \frac{4(1+\lambda)}{\lambda T^2} \left( \frac{z_\alpha + \Phi^{-1}(\pi)}{\bar{\theta}} \right)^2 \bm{J}^\top \left(\bm{C} + \lambda \bm{D}\right) \bm{J}.
\end{align*}

This provides the minimum sample size $ N $ required to achieve power $ \pi $ for the given effect size $ \bar{\theta} $ and covariance structure defined by $ \bm{C} $ and $ \bm{D} $.

\section{Proof of Theorem~\ref{thm4}} \label{Sec:app:Them4}

We have the estimator for the overall treatment effect $\hat{\bar{\theta}}$ defined as:
\[
\hat{\bar{\theta}} = \frac{1}{T K n_x n_y} \sum_{k=1}^K \sum_{t=1}^T \sum_{i=1}^{n_x} \sum_{j=1}^{n_y} \left[2I(x_{itk} < y_{jtk}) - 1 \right] = \frac{1}{T K n_x n_y} \sum_{j=1}^{n_y} \sum_{k=1}^K \sum_{t=1}^T \sum_{i=1}^{n_x} \left[2I(x_{itk} < y_{jtk}) - 1 \right].
\]

Define $\mathcal{Z}_{ijkt} = 2 I(x_{itk} < y_{jtk}) - 1$, which is a binary random variable taking values in $\{-1, 1\}$. The sum of interest is $S = \sum_{i=1}^{n_x} \sum_{j=1}^{n_y} \sum_{k=1}^K \sum_{t=1}^T \mathcal{Z}_{ijkt}$. The variables $\mathcal{Z}_{ijkt}$ exhibit dependence across time points $t$ and outcomes $k$ for a fixed subject pair $(i, j)$ due to repeated measures. However, we assume independence across different subject pairs $(i, j)$ and $(i', j')$ for $i\neq i'$ or $j\neq j'$. This assumption holds because observations from distinct patients are collected independently in clinical settings. The variance of $S$ can be expressed as:
\[
\text{Var}(S) = \sum_{i=1}^{n_x} \sum_{j=1}^{n_y} \sum_{k=1}^K \sum_{t=1}^T \text{Var}(\mathcal{Z}_{ijkt}) + \sum_{i=1}^{n_x} \sum_{j=1}^{n_y} \sum_{(k,t) \neq (k',t')} \text{Cov}(\mathcal{Z}_{ijkt}, \mathcal{Z}_{ijk't'}).
\]

To establish the asymptotic normality of $\hat{\bar{\theta}}$, we apply the Lyapunov Central Limit Theorem (CLT). For this, we verify the Lyapunov condition, which serves as an alternative to the Lindeberg condition. The Lyapunov condition states that for some $\delta > 0$, the following must hold:
\[
\lim_{n_x, n_y \to \infty} \frac{1}{\sigma_{n_x, n_y}^{2 + \delta}} \sum_{i=1}^{n_x} \sum_{j=1}^{n_y} \sum_{k=1}^K \sum_{t=1}^T E\left[ |\mathcal{Z}_{ijkt} - E[\mathcal{Z}_{ijkt}]|^{2 + \delta} \right] = 0,
\]
where $\sigma_{n_x, n_y}^2$ is the variance of $S$.

Since $\mathcal{Z}_{ijkt}$ takes values in $\{-1, 1\}$, we have $\mathcal{Z}_{ijkt}^2 = 1$. The variance of each $\mathcal{Z}_{ijkt}$ is $\text{Var}(\mathcal{Z}_{ijkt}) = 4p(1 - p)$, where $p = P(X_{tk} < Y_{tk})$. The higher moment $E\left[ |\mathcal{Z}_{ijkt} - (2p - 1)|^{2 + \delta} \right]$ is bounded, as $\mathcal{Z}_{ijkt}$ is symmetric and has only two possible values. Consequently, the sum inside the Lyapunov condition becomes:
\[
T K n_x n_y \cdot E\left[ |\mathcal{Z}_{ijkt} - (2p - 1)|^{2 + \delta} \right].
\]

Thus, the Lyapunov condition will be satisfied if the growth of the variance $\sigma_{n_x, n_y}^2$ outpaces the higher moments of $\mathcal{Z}_{ijkt}$ as $n_x$ and $n_y$ approach infinity.

The Lyapunov condition simplifies to verifying that the following expression tends to zero as $n_x, n_y \to \infty$:
\[
\frac{T K n_x n_y \cdot E\left[ |\mathcal{Z}_{ijkt} - (2p - 1)|^{2 + \delta} \right]}{\sigma_{n_x, n_y}^{2 + \delta}} \sim \frac{E\left[ |\mathcal{Z}_{ijkt} - (2p - 1)|^{2 + \delta} \right]}{\left[4p(1 - p)\right]^{1 + \delta/2}} \cdot \frac{1}{(T K n_x n_y)^{\delta/2}}.
\]

Since the higher-order moment $E\left[ |\mathcal{Z}_{ijkt} - (2p - 1)|^{2 + \delta} \right]$ is bounded due to $\mathcal{Z}_{ijkt}$ taking only two possible values, the key factor determining whether the Lyapunov condition holds is the growth rate of $\sigma_{n_x, n_y}^2$. As $\sigma_{n_x, n_y}^2$ grows proportionally with the sample size, the term $\frac{1}{(T K n_x n_y)^{\delta/2}}$ tends to zero as $n_x, n_y \to \infty$. This ensures that the Lyapunov condition is satisfied. Therefore, by the Lyapunov Central Limit Theorem, the normalized sum $\frac{S - E[S]}{\sigma_{n_x, n_y}}$ converges in distribution to a standard normal distribution: $
\frac{S - E[S]}{\sigma_{n_x, n_y}} \xrightarrow{d} \mathcal{N}(0, 1)$. This implies that $S$ is asymptotically normally distributed. Consequently, $\hat{\bar{\theta}}$ is also asymptotically normally distributed with mean $\bar{\theta}$ and variance given by:
\[
\text{Var}(\hat{\bar{\theta}}) = \frac{\text{Var}(S)}{T^2 K^2 n_x^2 n_y^2} = \frac{4p(1 - p)}{T K n_x n_y} + \frac{\sum_{\neq} \text{Cov}(\mathcal{Z}_{ijkt}, \mathcal{Z}_{i'j'k't'})}{T^2 K^2 n_x^2 n_y^2},
\]
where $\sum_{\neq}$ represents the sum of covariances across all $(k, t) \neq (k', t')$, formally written as $\sum_{\neq} = \sum_{i=1}^{n_x} \sum_{j=1}^{n_y} \sum_{(k, t) \neq (k', t')}$. Thus, the asymptotic normality of $\hat{\bar{\theta}}$ is established, which serves as a critical component in deriving the asymptotic distribution of the power estimate $\hat{\mathcal{P}}$.

The estimated power of the LRST is given by:
\[
\hat{\mathcal{P}} = \Phi \left( \frac{\hat{\bar{\theta}}}{\sqrt{\frac{4 \bm{J}^\top \hat{\bm{\Sigma}} \bm{J}}{N T^2}}} - z_{\alpha} \right),
\]
where $\Phi(\cdot)$ denotes the cumulative distribution function of the standard normal distribution, $\hat{\bar{\theta}}$ represents the estimated overall treatment effect, and $\bm{J}^\top \hat{\bm{\Sigma}} \bm{J}$ is a quadratic form involving the estimated covariance matrix $\hat{\bm{\Sigma}}$.

It has been established that $\hat{\bar{\theta}}$ is asymptotically normally distributed with mean $\bar{\theta}$ and variance $\sigma_{\theta}^2 = \frac{\text{Var}(S)}{T K n_x n_y}$. Similarly, $\bm{J}^\top \hat{\bm{\Sigma}} \bm{J}$ is asymptotically normal with mean $\bm{J}^\top \bm{\Sigma} \bm{J}$ and variance $\sigma^2_{\Sigma}$, where $\sigma^2_{\Sigma}$ is derived in Corollary~\ref{corro1}.

To derive the asymptotic distribution of $\hat{\mathcal{P}}$, consider the function:
\[
g(x, y) = \Phi \left( \frac{x}{\sqrt{\frac{4 y}{N T^2}}} - z_{\alpha} \right),
\]
where $x$ corresponds to $\hat{\bar{\theta}}$ and $y$ corresponds to $\bm{J}^\top \hat{\bm{\Sigma}} \bm{J}$. The function $g(x, y)$ describes the estimated power as a function of these two asymptotically normal random variables.

To apply the Delta Method, we compute the partial derivatives of $g(x, y)$ with respect to $x$ and $y$. The partial derivative with respect to $x$ is:
\[
\frac{\partial g}{\partial x} = \phi \left( \frac{x}{\sqrt{\frac{4 y}{N T^2}}} - z_{\alpha} \right) \cdot \frac{1}{\sqrt{\frac{4 y}{N T^2}}},
\]
where $\phi(\cdot)$ is the standard normal density function.

The partial derivative with respect to $y$ is:
\[
\frac{\partial g}{\partial y} = -\phi \left( \frac{x}{\sqrt{\frac{4 y}{N T^2}}} - z_{\alpha} \right) \cdot \frac{x}{2 y^{3/2}} \cdot \sqrt{\frac{N T^2}{4}}.
\]

Let $\xi = \frac{x}{\sqrt{\frac{4 y}{N T^2}}} - z_{\alpha}$ denote the standardized argument of the density function. These partial derivatives are essential for evaluating the asymptotic variance of $\hat{\mathcal{P}}$ in the subsequent derivations.

Let $\sigma_{\theta} = \frac{\text{Var}(S)}{T K n_x n_y}$ and $\sigma_{\Sigma} = \sigma_{\bm{J}^\top \bm{\Sigma} \bm{J}}$. Applying the Delta Method, we derive the asymptotic distribution of $\hat{\mathcal{P}} = g(\hat{\bar{\theta}}, \bm{J}^\top \hat{\bm{\Sigma}} \bm{J})$. Specifically, $\hat{\mathcal{P}}$ is asymptotically normally distributed as follows:
\[
\hat{\mathcal{P}} \xrightarrow{d} \mathcal{N} \left( g(\bar{\theta}, \bm{J}^\top \bm{\Sigma} \bm{J}), \, \text{Var}(\hat{\mathcal{P}}) \right),
\]
where the asymptotic variance $\text{Var}(\hat{\mathcal{P}})$ is given by:
\[
\text{Var}(\hat{\mathcal{P}}) = \left( \frac{\partial g}{\partial x} \right)^2 \sigma_{\theta}^2 + \left( \frac{\partial g}{\partial y} \right)^2 \sigma_{\Sigma}^2 + 2 \cdot \frac{\partial g}{\partial x} \cdot \frac{\partial g}{\partial y} \cdot \text{Cov}(\hat{\bar{\theta}}, \bm{J}^\top \hat{\bm{\Sigma}} \bm{J}),
\]
with $x = \hat{\bar{\theta}}$ and $y = \bm{J}^\top \hat{\bm{\Sigma}} \bm{J}$.

Substituting the expressions for the partial derivatives into the variance formula yields:
\[
\text{Var}(\hat{\mathcal{P}}) = \phi^2 \left( \frac{\bar{\theta}}{\sqrt{\frac{4 \bm{J}^\top \bm{\Sigma} \bm{J}}{N T^2}}} - z_{\alpha} \right)
\left( 
\frac{\sigma_{\theta}^2}{\frac{4 \bm{J}^\top \bm{\Sigma} \bm{J}}{N T^2}} 
+ \frac{\bar{\theta}^2 \sigma_{\Sigma}^2}{4 (\bm{J}^\top \bm{\Sigma} \bm{J})^3} \cdot \frac{N T^2}{4} 
- 2 \cdot \frac{\sigma_{\theta} \bar{\theta} \, \text{Cov}(\hat{\bar{\theta}}, \bm{J}^\top \hat{\bm{\Sigma}} \bm{J})}{2 (\bm{J}^\top \bm{\Sigma} \bm{J})^{3/2}} \cdot \frac{N T^2}{4} 
\right).
\]

Analyzing each term:

The first term simplifies to:
\[
\frac{\sigma_{\theta}^2}{\frac{4 \bm{J}^\top \bm{\Sigma} \bm{J}}{N T^2}} 
= \frac{\text{Var}(S) N T^2}{4 T^2 K^2 n_x^2 n_y^2 \bm{J}^\top \bm{\Sigma} \bm{J}} 
= \frac{T^2}{4 \bm{J}^\top \bm{\Sigma} \bm{J}} \left\{ \frac{4 p(1-p) N}{T K n_x n_y} + \frac{N \sum_{\neq} \text{Cov}(\mathcal{Z}_{ijkt}, \mathcal{Z}_{ijk't'})}{T^2 K^2 n_x^2 n_y^2} \right\} \rightarrow 0.
\]

The second term becomes:
\[
\frac{\bar{\theta}^2 \sigma_{\Sigma}^2}{4 (\bm{J}^\top \bm{\Sigma} \bm{J})^3} \cdot \frac{N T^2}{4} 
= \frac{\bar{\theta}^2 N T^2}{16 (\bm{J}^\top \bm{\Sigma} \bm{J})^3} 
\left[ 
\frac{1}{K^4} \sum_{t_1 \neq t_2} \sum_{k_1, k_2 = 1}^K \mathcal{G}_{t_1 k_1 t_2 k_2} 
+ \frac{1}{K^4} \sum_{t=1}^T \left( \sum_{k_1 \neq k_2} \mathcal{G}_{t k_1 t k_2} + \sum_{k=1}^K \mathcal{H}_{t k t k} \right)
\right].
\]

The final term, without the covariance term explicitly, simplifies to:
\[
\frac{\sigma_{\theta} \bar{\theta} N T^2}{4 (\bm{J}^\top \bm{\Sigma} \bm{J})^{3/2}} 
= \frac{\bar{\theta} T^2}{4 (\bm{J}^\top \bm{\Sigma} \bm{J})^{3/2}} 
\sqrt{ 
\frac{4 p(1-p) N^2}{T K n_x n_y} 
+ \frac{N^2 \sum_{\neq} \text{Cov}(\mathcal{Z}_{ijkt}, \mathcal{Z}_{i'j'k't'})}{T^2 K^2 n_x^2 n_y^2} 
}.
\]

As $n_x, n_y \to \infty$, this term converges to:
\[
\frac{\bar{\theta} \left( \lambda + \frac{1}{\lambda} + 2 \right) T^2}{4 (\bm{J}^\top \bm{\Sigma} \bm{J})^{3/2}} 
\sqrt{ 
\frac{4 p(1-p)}{T K} 
+ \frac{\text{Cov}(\mathcal{Z}_{11kt}, \mathcal{Z}_{11k't'})}{T^2 K^2} 
}.
\]

The remaining task is to derive $ \text{Cov}(\hat{\bar{\theta}}, \bm{J}^\top \hat{\bm{\Sigma}} \bm{J}) $.

\begin{align*}
    \text{Cov}\left(\hat{\bar{\theta}}, \bm{J}^\top \hat{\bm{\Sigma}} \bm{J} \right) &= \text{Cov} \left[ \frac{1}{T K n_x n_y} \sum_{j=1}^{n_y} \sum_{k=1}^K \sum_{t=1}^T \sum_{i=1}^{n_x} \left(2I(x_{itk} < y_{jtk}) - 1 \right), \sum_{t_1=1}^T \sum_{t_2=1}^T \hat{\Sigma}_{t_1,t_2} \right] \\
    &= \text{Cov} \left\{ \frac{1}{T K n_x n_y} \sum_{j=1}^{n_y} \sum_{k=1}^K \sum_{t=1}^T \sum_{i=1}^{n_x} \left(2I(x_{itk} < y_{jtk}) - 1 \right), \right. \\
    &\quad \left. \frac{1}{K^2} \sum_{t_1, t_2=1}^T \sum_{k_1, k_2=1}^K \left[\left(1 + \frac{1}{\lambda}\right) \hat{c}_{t_1k_1t_2k_2} + (1 + \lambda) \hat{d}_{t_1k_1t_2k_2} \right] \right\}.
\end{align*}

To compute this covariance, we consider the components $ \text{Cov}\left( 2I(x_{itk} < y_{jtk}) - 1, \hat{c}_{t_1k_1t_2k_2} \right) $ and \\$ \text{Cov}\left( 2I(x_{itk} < y_{jtk}) - 1, \hat{d}_{t_1k_1t_2k_2} \right) $.

For the $ \hat{c}_{t_1k_1t_2k_2} $ term:

\begin{align*}
    \text{Cov} \left[ I(x_{itk} < y_{jtk}), \hat{c}_{t_1k_1t_2k_2} \right] 
    &= \text{Cov} \left\{ I(x_{itk} < y_{jtk}), \frac{1}{n_x} \sum_{u=1}^{n_x} \left[ \frac{1}{n_y} \sum_{v=1}^{n_y} I(y_{v t_1 k_1} < x_{u t_1 k_1}) - \frac{1 - \hat{\theta}_{t_1 k_1}}{2} \right] \right. \\
    & \quad \left. \times \left[ \frac{1}{n_y} \sum_{v=1}^{n_y} I(y_{v t_2 k_2} < x_{u t_2 k_2}) - \frac{1 - \hat{\theta}_{t_2 k_2}}{2} \right] \right\} \\
    &= \frac{1}{n_x n_y} \, \text{Cov} \left[ I(x_{itk} < y_{jtk}), I(y_{j t_1 k_1} < x_{i t_1 k_1} \, \& \, y_{j t_2 k_2} < x_{i t_2 k_2}) \right].
\end{align*}

Similarly, for the $ \hat{d}_{t_1k_1t_2k_2} $ term:

\[
\text{Cov} \left[ I(x_{itk} < y_{jtk}), \hat{d}_{t_1k_1t_2k_2} \right] 
= \frac{1}{n_x n_y} \, \text{Cov} \left[ I(x_{itk} < y_{jtk}), I(y_{j t_1 k_1} > x_{i t_1 k_1} \, \& \, y_{j t_2 k_2} > x_{i t_2 k_2}) \right].
\]

Combining these results, we find:

\[
\text{Cov} \left( \hat{\bar{\theta}}, \bm{J}^\top \hat{\bm{\Sigma}} \bm{J} \right) 
= \frac{1}{T K^3 n_x n_y} \sum_{i=1}^{n_x} \sum_{j=1}^{n_y} \sum_{k, k_1, k_2=1}^K \sum_{t, t_1, t_2=1}^T O \left( \frac{1}{n_x n_y} \right) \rightarrow 0.
\]

Thus, the covariance between $\hat{\bar{\theta}}$ and $\bm{J}^\top \hat{\bm{\Sigma}} \bm{J}$ asymptotically approaches zero as the sample sizes $n_x$ and $n_y$ grow large ($n_x, n_y \to \infty$). The above derivation demonstrates that the only non-vanishing term in the asymptotic variance of $\hat{\mathcal{P}}$ is the second term:

\[
\frac{\bar{\theta}^2 N T^2}{16 (\bm{J}^\top \bm{\Sigma} \bm{J})^3} \left[ \frac{1}{K^4} \sum_{\substack{t_1 \neq t_2 \\ t_1, t_2 = 1}}^T \sum_{k_1, k_2 = 1}^K \mathcal{G}_{t_1k_1t_2k_2} + \frac{1}{K^4} \sum_{t=1}^T \left( \sum_{\substack{k_1 \neq k_2 \\ k_1, k_2=1}}^K \mathcal{G}_{t k_1 t k_2} + \sum_{k=1}^K \mathcal{H}_{t k t k} \right) \right].
\]

Therefore, the final asymptotic distribution of the power estimate \( \hat{\mathcal{P}} \) is: $\hat{\mathcal{P}} \xrightarrow{d} \mathcal{N}\left( \mathcal{P}, \text{Var}(\hat{\mathcal{P}}) \right)$, where $\mathcal{P}$ denotes the theoretical power, and the asymptotic variance is given by:

\[
\text{Var}\left(\hat{\mathcal{P}}\right) = \phi^2\left( \frac{\bar{\theta}}{\sqrt{\frac{4 \bm{J}^\top \bm{\Sigma} \bm{J}}{N T^2}}} - z_{\alpha} \right) 
\cdot \frac{\bar{\theta}^2 N T^2}{16 (\bm{J}^\top \bm{\Sigma} \bm{J})^3} 
\left[ \frac{1}{K^4} \sum_{\substack{t_1 \neq t_2 \\ t_1, t_2 = 1}}^T \sum_{k_1, k_2 = 1}^K \mathcal{G}_{t_1k_1t_2k_2} 
+ \frac{1}{K^4} \sum_{t=1}^T \left( \sum_{\substack{k_1 \neq k_2 \\ k_1, k_2=1}}^K \mathcal{G}_{t k_1 t k_2} + \sum_{k=1}^K \mathcal{H}_{t k t k} \right) \right].
\]

Here, $\phi(\cdot)$ represents the probability density function of the standard normal distribution, $\bar{\theta}$ is the estimated overall treatment effect, and $\bm{J}^\top \bm{\Sigma} \bm{J}$ is the quadratic form involving the covariance matrix $\bm{\Sigma}$. The terms $\mathcal{G}_{t_1k_1t_2k_2}$ and $\mathcal{H}_{tktk}$ capture the variance components across time points and within specific outcomes, respectively.







\end{document}